\documentclass{ws-book9x6}
\usepackage{ws-book-thm}  
\usepackage{ws-book-har}   
\usepackage[pdfpagelabels=false,colorlinks=true,allcolors=black]{hyperref}  
\usepackage{slashed}
\usepackage[normalem]{ulem}
\usepackage{subfigure}
\usepackage{sidecap}

\providecommand{\pasa}{Publ.\ Astron.\ Soc.\ Austr.}
\providecommand{\aap}{Astron.\ Astrophys.}
\providecommand{\apjs}{Astrophys.\ J.\ Suppl.}
\providecommand{\apjl}{Astrophys.\ J.\ Lett.}
\providecommand{\apj}{Astrophys.\ J.}
\providecommand{\physrep}{Phys.\ Rep.}
\providecommand{\mnras}{Mon.\ Not.\ Roy.\ Astron.\ Soc.}

\providecommand{\rmp}{Rev.\ Mod.\ Phys.}
\providecommand{\prc}{Phys.\ Rev.\ C}
\providecommand{\prd}{Phys.\ Rev.\ D}
\providecommand{\prx}{Phys.\ Rev.\ X}

\providecommand{\prl}{Phys.\ Rev.\ Lett.}

\providecommand{\nat}{Nature}

\title{Astrophysics in the XXI Century with Compact Stars}
\date{\today}

\begin{document}

\chapter[Effects of phase transition on supernova explosions, compact stars and mergers]{Effects of a strong phase transition on supernova explosions, compact stars and their mergers}
\label{ch_bbf}

\author[
Andreas Bauswein, David Blaschke and Tobias Fischer]
{Andreas Bauswein$^{1}$, David B. Blaschke$^{2}$ and Tobias Fischer$^{2}$,}
\address{
$^{1}$~GSI Helmholtzzentrum f{\"u}r Schwerionenforschung GmbH, Darmstadt, Germany
$^{2}$~Institute of Theoretical Physics, University of Wroclaw, Poland
\\ 
}


\section*{Abstract}
\label{abstract}
We outline a theoretical approach supporting strong phase transitions from normal nuclear matter to the deconfined quark-gluon plasma, in the equation of state (EOS) for compact star matter. Implications of this hypothesis are discussed for astrophysical applications. Special emphasis is devoted to potentially detectable signatures, which can be directly related with the occurrence of a sufficiently strong phase transition. Therefore, simulations of core-collapse supernovae and binary compact star mergers are considered, including the subsequent emission of gravitational waves and, in the case of supernova, in addition the neutrinos play the role of messengers. 

\section{Introduction}
\label{sec1.1}
The state of matter at the interior of core-collapse supernovae and binary neutron star mergers can reach extreme conditions, in terms of temperatures up to several $10^{11}$~K \footnote{$1.16\times10^{10}~\rm K\simeq 1$~MeV}, baryon densities in excess of normal nuclear saturation density and large isospin asymmetry. The latter is given by the proton fraction (in general charge fraction) $Y_p$, where $Y_p=0$ for neutron matter (no net charges are present) and $Y_p=0.5$ for isospin symmetric matter. The EOS for simulations of supernovae and binary neutron star mergers must hence cover such an extended three-dimensional domain, where presently first-principle nuclear matter calculations are not available. Instead, model EOS are being developed for supernova studies or simulations of neutron star mergers \citep[for a comprehensive review about the EOS in supernova studies, c.f.][]{Fischer:2017}. These combine several domains with different degrees of freedom, e.g., heavy nuclei at low temperatures, inhomogeneous nuclear matter composed of light and heavy nuclei together with unbound nucleons, and homogeneous matter at high temperatures and densities. In this chapter we reflect on the role of the EOS in core-collapse supernovae and during binary neutron star mergers, with particular focus on the recent finding about the presence of a phase transition from normal hadronic matter to the quark-gluon plasma. 

In the case of core-collapse supernova studies, accurate neutrino transport is essential for the prediction of the neutrino signal from these explosive events, as was observed from SN1987A \citep[cf.][]{Bionta:1987qt,Hirata:1988ad}. This observation is until now a benchmark. It confirmed that neutrinos from the next galactic event will become {\em the} observable signal, from which we will be able to learn not only details about these stellar explosions but also about the state of matter at the supernova interior. In addition to neutrinos, also gravitational waves are emitted from the interior of core-collapse supernovae. The {\em classical} emission is associated with signal that originates from the core bounce of moderately up to rapidly rotating stellar cores collapse \citep[c.f.][]{Ott:2004,Ott:2006,Kotake:2006,Dimmelmeier:2007,Dimmelmeier:2008,Scheidegger:2008}, which, however, seems difficult to be detectable. In general, the SN GW signal does not obey any template character. For example, it depends sensitively on the details of the radiation hydrodynamics models used to provide GW signals form simulations and on the EOS, as was demonstrated by \citet{Scheidegger:2010}. An alternative idea for the detection prospects of gravitational wave from the long-term post-bounce evolution has been proposed by \citet{Murphy09}. The foundation of this is the excitation of a few dominant fluid modes, which, in turn, give rise to the emission of GWs. This idea has been further explored and a connection could be established between the neutrino and GW emission characteristics using sophisticated general relativistic three-dimensional SN simulations by \citet{Kuroda:2017}. Nevertheless, the lack of any SN GW measurement, for any of the observing runs of aLIGO/VIRGO,  demonstrate the complication of the problem and point to the need of improvements, both, of the GW detectors in the kHZ range as well as on the data analysis.

Gravitational waves are also key observable from binary neutron star mergers. Since the equation of state of neutron star matter determines the stellar structure and the dynamics of a neutron star merger, gravitational waves carry information about the properties of high-density matter. This in particular includes the prospect to learn about the presence of deconfined quark matter in compact stars from the observations of gravitational waves produced by neutron star mergers.

While galactic supernova explosions and neutron star mergers provide the ultimate sites for testing the location and characteristics of the deconfinement transition in the QCD phase diagram at high baryon densities and low temperatures, inaccessible to heavy-ion collision experiments, they are very rare events. Since SN1987A occurred we have not observed a supernova neutrino signal and since GW170817, there was no binary neutron star merger detected which would allow the extraction of the neutron star compactness  from the measurement of its tidal deformability. Up to now, we have no detection of the postmerger gravitational wave signal.  Therefore, compact stars which are observed as pulsars in different astrophysical environments, remain excellent tools to calibrate our knowledge of the dense matter equation of state albeit at zero temperature. 


We note that the ab-initio simulations of lattice QCD cannot make at present any predictions for the thermodynamics in the QCD phase diagram at finite and even high baryon density, due to the sign problem;
see \cite{Muroya:2003qs} for an introductory review on this issue.
Present techniques of lattice quantum field theory for evaluating 
the QCD partition function are applicable only for low baryochemical potentials $\mu_B/T\lesssim 2$.
It is found that the QCD transition is a chiral crossover with 
a pseudo-critical temperature $T_c(0)=156.5\pm 1.5$ MeV at $\mu_B=0$
\cite{HotQCD:2018pds}.
There is no indication for a critical endpoint (CEP) for $\mu_B/T\lesssim 2$ and $T>0.9~T_c(0)$  \cite{Bazavov:2017dus}. 
From recent lattice QCD simulations at very low quark masses and their extrapolation to the chiral limit it has been concluded that the chiral phase transition temperature is $T_c^{(0)}=132^{+3}_{-6}$ MeV. For the temperature of the CEP (if that exists at all) has to hold $T_{\rm CEP} < T_c^{(0)} $ \cite{HotQCD:2019xnw}.
\\
At low baryon chemical potentials and high temperatures the QCD EOS measured in lattice QCD simulations is well confirmed by a Bayesian analysis of high-statistics data obtained by heavy-ion collision (HIC) experiments at the Relativistic Heavy-Ion Collider and at the Large Hadron Collider \cite{Pratt:2015zsa}.
It is the hope that in systematic beam energy scan programs of HIC experiments at lower collision energies ($\sqrt{s_{NN}}=3 \dots 11$ GeV being the relevant energy range in the nucleon-nucleon center of mass system of the collision), similar analyses of the high-statistics data would reveal indications of the CEP as a landmark in the QCD phase diagram. 
This hope, however, has not been fulfilled yet. For a recent review, see \cite{An:2021wof}.   
The problem has been clearly elucidated in the work by 
\citet{Senger:2021dot}, where he shows that when the temperature of the chemical freezeout (dilepton spectrum) should fulfill the above lattice QCD constraint for $T_{\rm CEP}$, the collision energy 
has to be below $\sqrt{s_{NN}}= 6 (3)$ GeV. 
In order to reach the mixed phase of a first-order phase transition with the dynamical trajectories of HIC simulations in the QCD phase diagram, a collision energy of at least $\sqrt{s_{NN}}= 5$ GeV is required \cite{Arsene:2006vf}.
Following this line of argument, even if a CEP exists, its temperature is too low to be reached in HIC that could probe the QCD phase transition.

Therefore, in order to investigate the hadron-to-quark matter phase transition at low temperatures,
one has to rely on observations of compact stars in their multi-messenger astrophysical context: as proto-neutron stars in supernova explosions, as pulsars in isolation and in binaries as well as in their final stages of their life, in the inspiral, merger and postmerger phases of binary neutron star mergers.

In this contribution, we review several, recent aspects of the role that a strong QCD phase transition can play in these three. We discuss the birth of compact stars in core-collapse supernova events as well as the astrophysics of binary neutron star merger events.

\section{Hybrid neutron star formation in core-collapse supernova explosions} 
\label{sec1.3}
Proto-neutron stars (PNS) are the central, compact objects of core-collapse supernovae. They are the transitional state of the initial collapse of the stellar core of massive stars and the remnant neutron star. The stellar core collapse is triggered due to the photodisintegration of heavy, iron-group nuclei and due to electron captures on protons bound in these nuclei. Consequently, the density and the temperature increase continuously during the collapsing stellar core, which simultaneously deleptonises. When nuclear saturation density is reached in the center, the short range repulsive nuclear force balances gravity such that the core bounces back with the formation of a hydrodynamics shock wave. The latter propagates quickly out of the stellar core and stalls at a radius of around 100--200~km due to the continuous photodisintegration of infalling heavy nuclei from above and the launch of the $\nu_e$-burst associated with the shock propagation across the neutrino sphere of last scattering. As a result, the dynamic shock turns into an accretion front.

The supernova explosions are related with the revival of the stalled bounce shock wave and the subsequent ejection of the stellar mantle that surrounds the PNS. It has long been considered via the liberation of energy from the PNS interior to a thin layer of accumulated material at the PNS surface \citep[for reviews about core collapse supernovae, see][]{Janka:2007PhR,Mirizzi:2016}. 
Several scenarios have been explored in the literature. The  magneto-rotational mechanism of \citet{LeBlanc:1970kg} \citep[for recent works, c.f.,][]{Takiwaki:2007sf,Winteler:2012,Moesta:2014,Moesta:2015}, considers the required energy through the winding up of the magentic field stress in the presence of initially high magnetic field and rapidly rotating stellar cores. Another scenario is the neutrino-heating mechanism of \citet{Bethe:1985ux}. With angle and energy dependent neutrino transport methods once can demonstrate the development of a gain region, below which neutrinos dominantly decouple from matter and hence establish net cooling, while above neutrinos are still dominantly trapped and hence net heating is established. 
This mechanism has been demonstrated to lead to supernova explosions for a variety of massive progenitor stars 
\citep[cf.][]{Mueller:2012a,Takiwaki:2012,Bruenn:2013,Melson:2015,Lentz:2015}. 
However, in the framework of multi-dimensional simulations accurate Boltzmann neutrino transport cannot be employed due to the current computational limitations. Instead, approximate neutrino transport schemes are commonly used, whose range of applicability is being debated currently \citep[cf.][]{Sumiyoshi:2018}. 

\subsection{Simulations of core-collapse supernovae with hadron-quark matter hybrid EOS}
Besides the aforementioned scenarios for the onset of the supernova explosion, another mechanism was discovered by \citet{Sagert:2008ka}, which is due to the occurance of a phase transition at high densities from hadronic matter to the quark gluon plasma. The latter was being treated within the simple but powerful bag model. During the phase transition a large amount of latent heat is released in a highly dynamical fashion, which in turn triggers the onset of the supernova explosion even in spherically symmetric simulations. Moreover, it leaves an observable millisecond burst in the neutrino signal, as was first demonstrated by \citet{Dasgupta:2009yj} for the currently operating Super-Kamiokande neutrino detector. 
These milestones demonstrate the sensitivity of EOS uncertainties related to our understanding of core-collapse supernovae and underline the need of a more elaborate understanding of the EOS at high densities including better constraints in particular. 

\begin{figure*}[t!]
\subfigure[~$T=0$]{
\includegraphics[width=0.995\columnwidth]{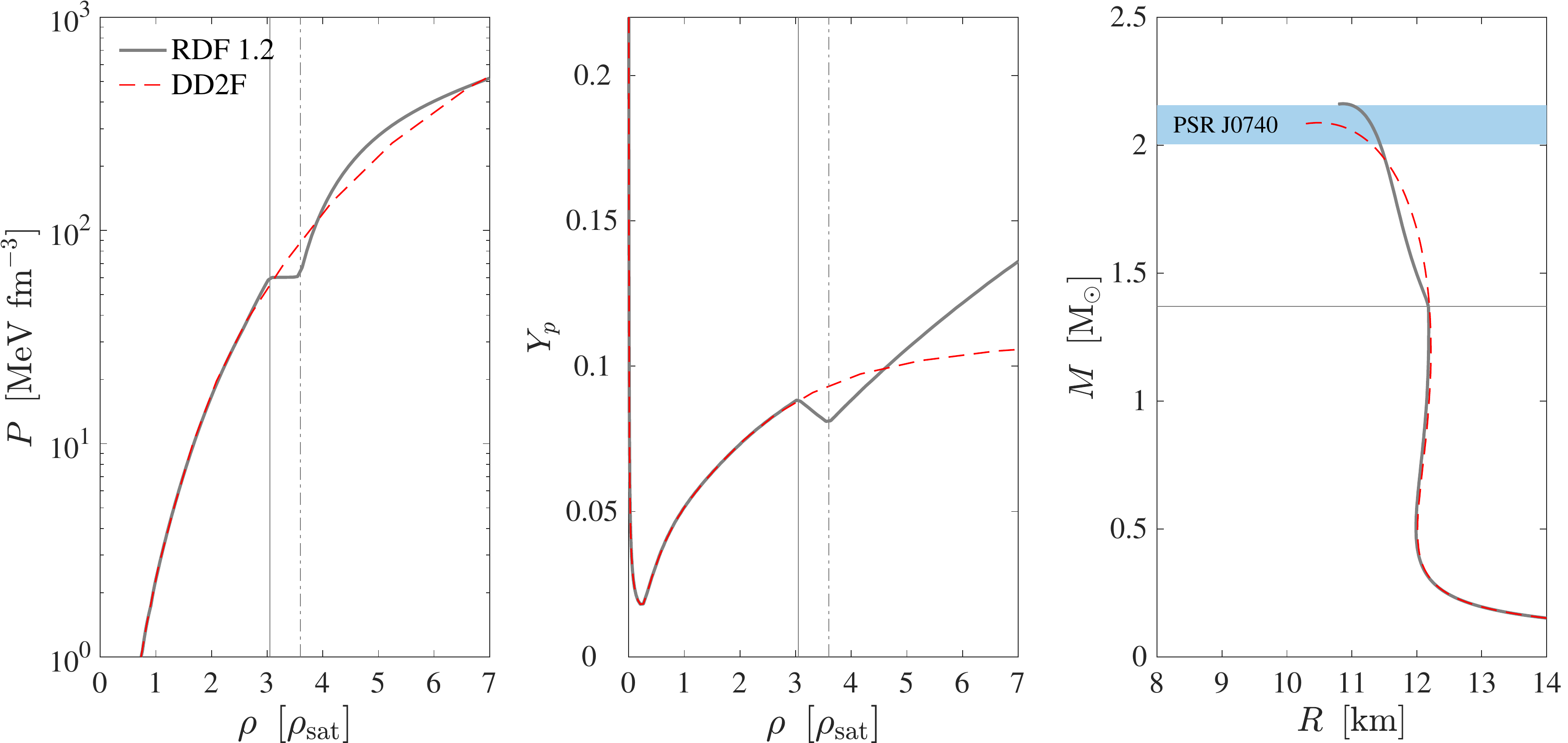}
\label{fig:sn_eos_a}
} 
\\
\subfigure[~$s=3~k_{\rm B}$]{\includegraphics[width=0.995\columnwidth]{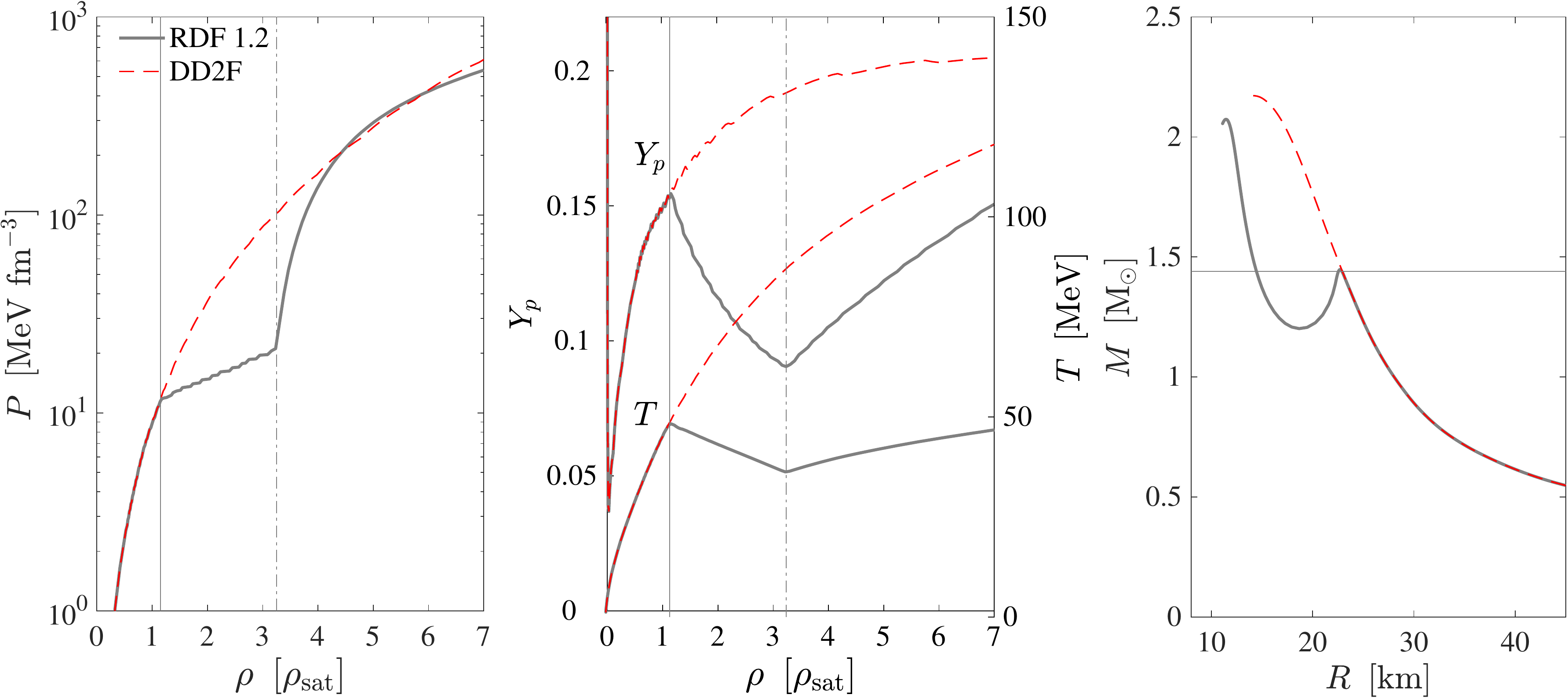}
\label{fig:sn_eos_b}
}
\caption{\citep[Figure adopted from][]{Fischer:2021} Equation of state assuming  $\beta$-equilibrium, showing the total pressure, $P$, the charge fraction, $Y_p$ and the temperature $T$, as a function of the baryon density, $\rho$, as well as the corresponding mass-radius relations, $M$--$R$, comparing the RDF~1.2 hadron-quark hybrid model (solid grey lines) and the DD2F reference hadronic model (red dashed lines).
}
\label{fig:sn_eos}
\end{figure*}

\begin{figure*}[t!]
\centering
\includegraphics[width=\columnwidth]{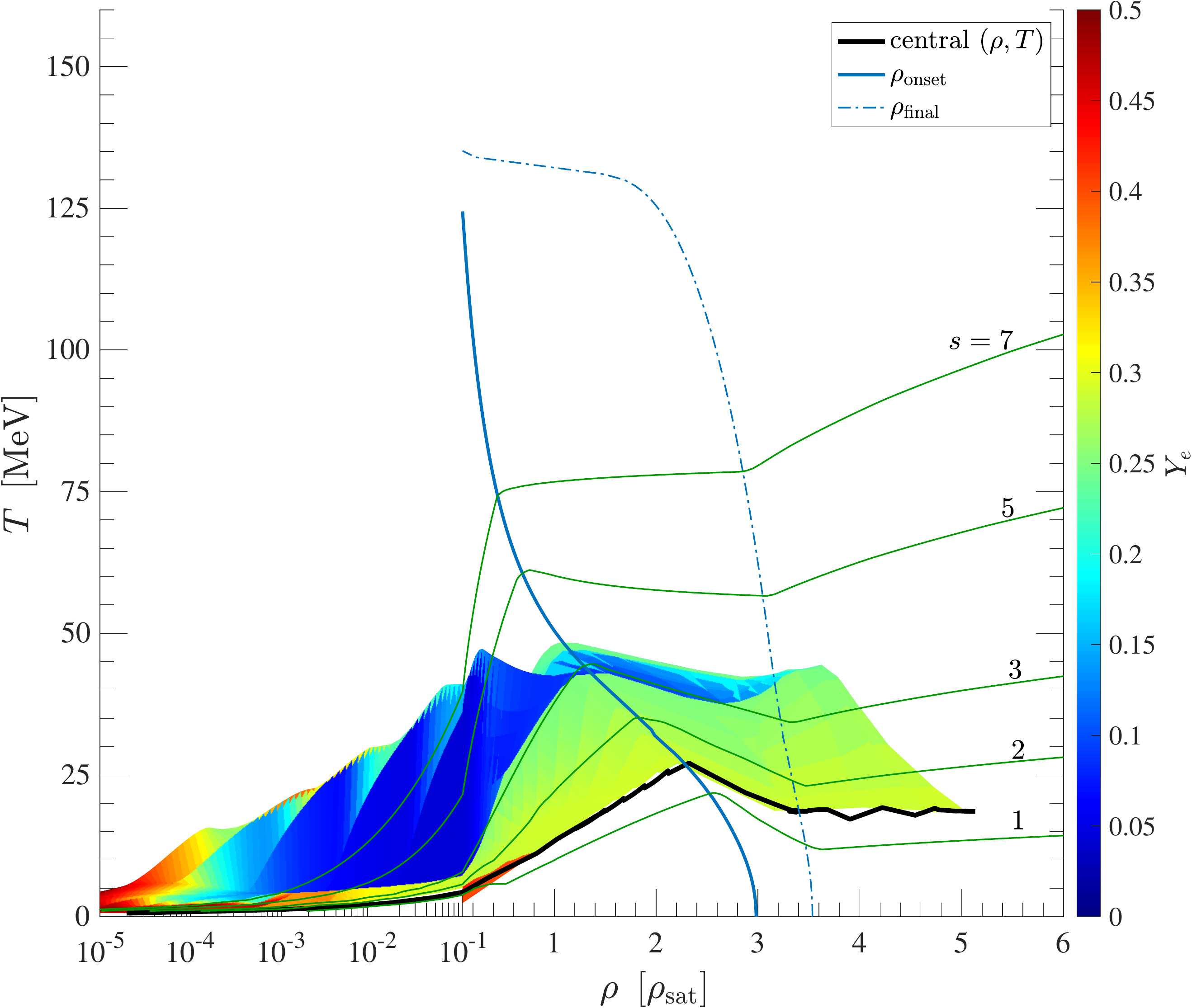}
\caption{\label{fig:sn_phasediagram} \citep[Figure adopted from][]{Fischer:2021} Supernova temperature--density region occupied in simulations that feature a hadron-quark phase transition, where the color-coding is due to the electron abundance. Shown are also the hadron-quark matter phase boundaries for the onset of quark matter, $\rho_{\rm onset}$, and for reaching the pure quark-matter phase, $\rho_{\rm final}$, for the DD2F-RDF~1.2 hybrid EOS, as well as curves of constant entropy (green solid lines) together with the central supernova trajectory (thick solid black line). 
}
\end{figure*}

\begin{figure*}[t!]
\centering
\includegraphics[width=0.9\columnwidth]{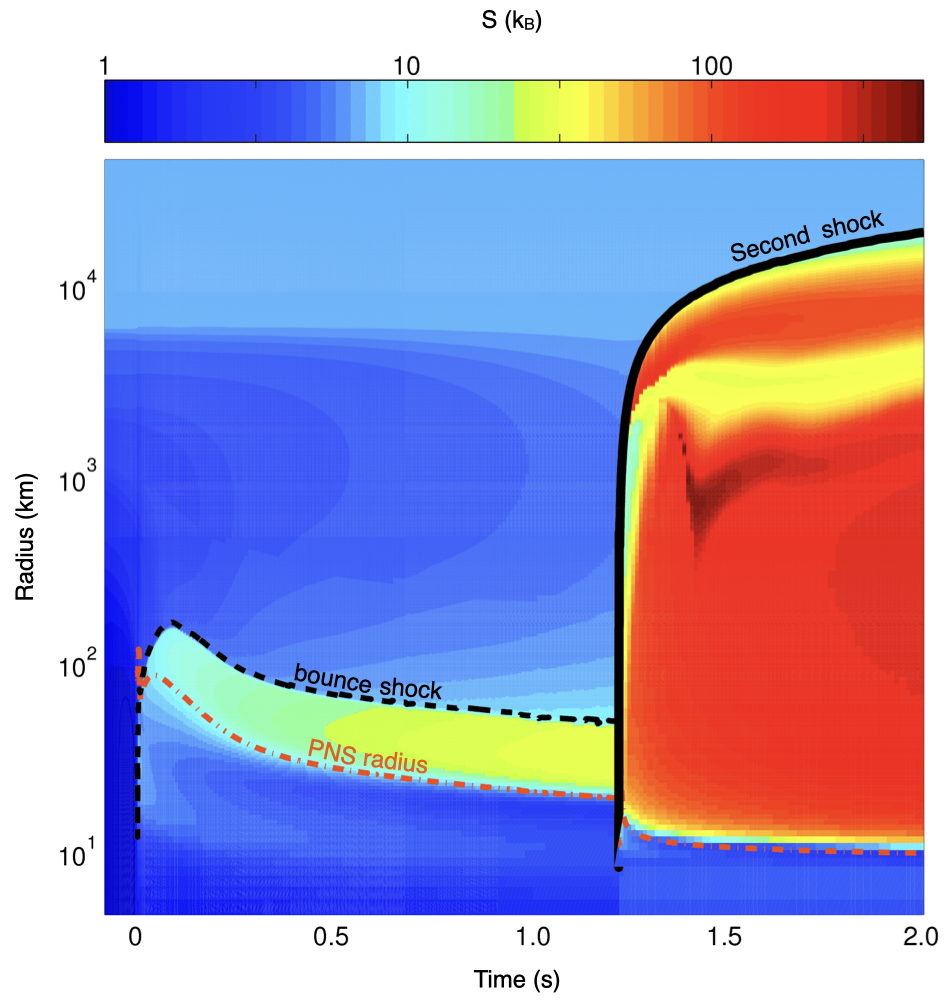}
\caption{\label{fig:sn_shellplot} \citep[Figure adopted from][]{Fischer:2018} Post-bounce evolution of the supernova launched from a 50~M$_\odot$ progenitor where the DD2F-RDF~1.2 hybrid EOS has been employed with the phase transition at about 1.2~s, where the entropy per particle being the color-code. Shown are the PNS radius, the bounce shock and the second shock, which forms as a direct consequence of the first-order hadron-quark phase transition.}
\end{figure*}

\begin{SCfigure*}
\includegraphics[width=0.725\columnwidth]{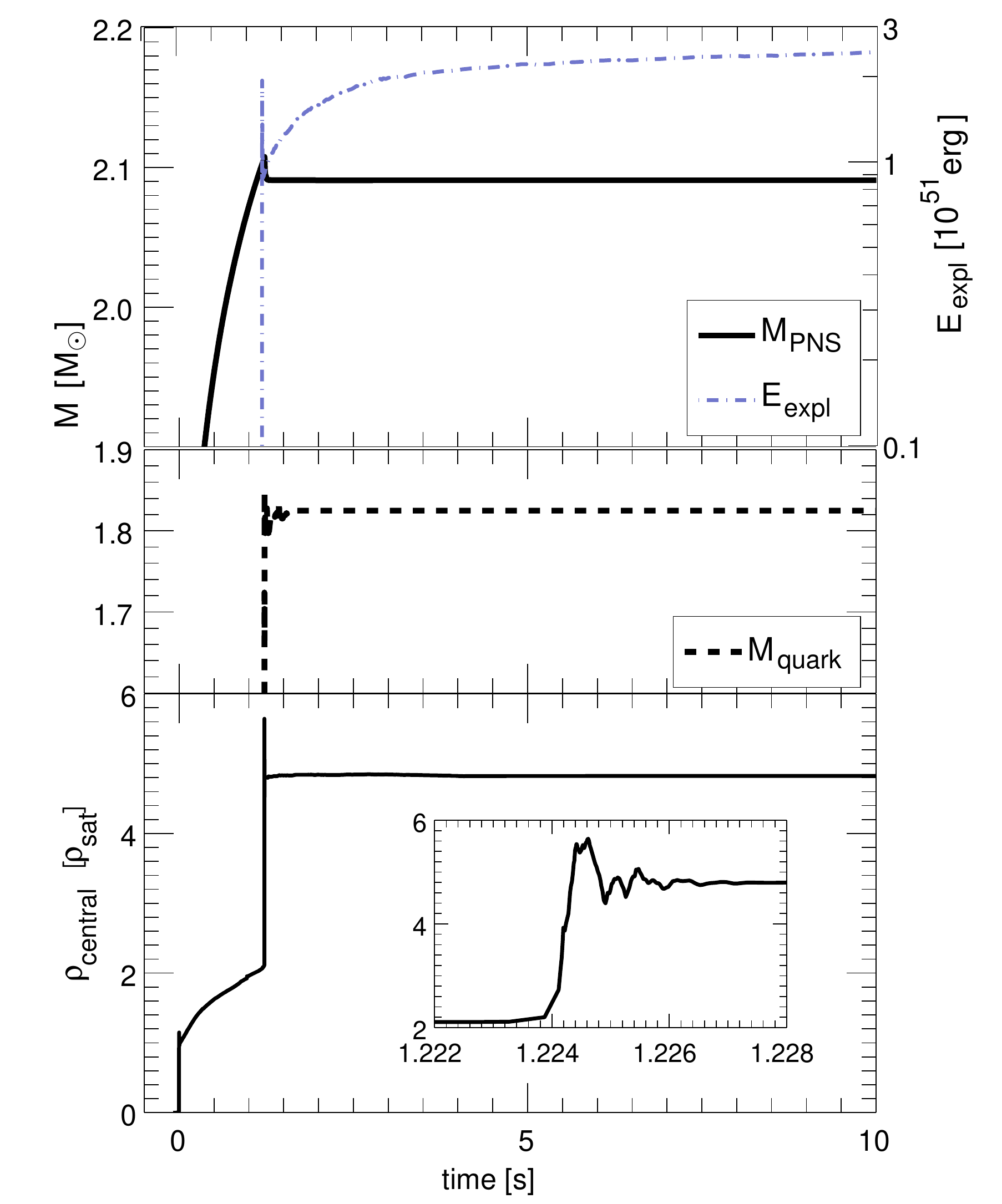}
\caption{\label{fig:sn_evol} \citep[Figure adopted from][]{Fischer:2018} Post-bounce supernova evolution of a 50~M$_\odot$ progenitor star that takes into account the hadron-quark matter phase transition, which occurs in this model after at about 1.2~s. The figure shows the explosion energy, $E_{\rm expl}$, the mass enclosed inside the PNS, $M_{\rm PNS}$, as well as the quark-matter core, $M_{\rm quark}$, and the central density, $\rho_{\rm central}$. }
\end{SCfigure*}

For example, there is one problem concerning the hybrid EOS used in \citet{Sagert:2008ka}, namely that it does not fulfill the constraint which arises from the observations of pulsars with masses as high as 2~M$_\odot$, demonstrated, e.g., by \citet{Antoniadis:2013,Fonseca:2016,Cromartie:2020NatAs,Fonseca:2021wxt}. This well-known problem of bag-model like hybrid EOS has been overcome by the inclusion of linear repulsive vector interaction, such as the VBAG model of \citet{Klaehn:2015}, similar to effective nuclear mean-field models. Higher-order repulsive interactions were considered in \citet{Benic:2015} and \citet{Kaltenborn:2017}, in light of the existence of a third family of compact stars as well as the so-called twin phenomenon where two families of compact stars exist with nearly the same mass but different radii. The corresponding RDF~1.2 hybrid EOS is shown in Fig.~\ref{fig:sn_eos}, for $\beta$-equilibrium and zero temperature conditions in Fig.~\ref{fig:sn_eos_a}, together with the mass-radius relation where the horizontal blue band marks the present maximum neutron star constraint of $2.08\pm 0.07$M$_\odot$ from the measurement of the pulsar PSR~J0740+6620 \citep[][]{Antoniadis:2013,Fonseca:2021wxt}. The vertical lines  in Fig.~\ref{fig:sn_eos} indicate the onset of the phase transition (solid) and reaching the pure quark matter phase (dash-dotted), which is due to Maxwell's condition for the first-order phase transition construction. In between is the region of thermodynamic instability. The need of data in between the onset of quark matter and reaching the pure quark matter phase, for hydrodynamics simulations, results in a hadron-quark mixed phase where the pressure gradient is flat in the zero-temperature case (see fig.~\ref{fig:sn_eos_a}). It is interesting to note that at finite entropy, examined at the example of $s=3~k_{\rm B}$, which is a representative value for the PNS interior, the phase boundaries are located at lower densities and the phase transition region widens (see Fig.~\ref{fig:sn_eos_b}). This has important consequences for the stability of PNS, in connection with cold hybrid stars, for which the onset mass increases but the maximum mass decreases \citep[a detailed discussion about the subject can be found in][]{Hempel:2016}.

To support the temperature dependence of the phase boundaries, $\rho_{\rm onset}$ and $\rho_{\rm final}$, Fig.~\ref{fig:sn_phasediagram} shows qualitative results for the DD2F-RDF~1.2 hadron-quark model EOS, together with curves of constant entropy per particle of $s=1-7~k_{\rm B}$. There is another important aspect related to the finite temperature dependence on the phase transition construction, the EOS softens substantially in the hadron-quark coexistence region (see Fig.~\ref{fig:sn_eos_b}). During the post-bounce evolution, the central PNS grows not only in mass but also the central density increases continuously, due to the continuous mass accretion. Once the central density reaches $\rho_{\rm onset}$ the PNS becomes unstable. The initial contraction proceeds into an adiabatic collapse, which halts when a sufficient amount of the PNS interior reaches the pure quark matter phase, where the EOS stiffens again. In the RDF EOS class this is additionally supported due to the vector interactions. As a consequence, an additional strong hydrodynamics shock wave is formed at the phase boundaries, which propagates quickly to increasingly larger radii (see the thick solid black line in Fig.~\ref{fig:sn_shellplot}). This second shock wave takes over the bounce shock, which triggers the onset of the supernova explosion. Afterwards, on the timescale of several hundreds of milliseconds, the central PNS, now featuring a massive core composed of deconfined quark matter (see Fig.~\ref{fig:sn_evol}), settles into a quasi-stationary state, i.e. the central density saturates at about five times the saturation density (see Fig.~\ref{fig:sn_evol}). 

The classical picture of core-collapse supernova phenomenology, originally proposed by \citet{Woosley:2002zz}, considers the high-mass progenitor contributions to the failed explosion branch, which yield the formation of black holes instead. It becomes evident that this picture might be incomplete when taking into account a supernova explosion scenario, driven by the hadron-quark phase transition. However, this cannot account for the bulk part of all massive star explosions. It affects mainly the high-mass end of massive stars, due to the rather high central densities required for the phase transition to occur, which are not reached for low- and intermediate-mass progenitors on the order of $\leq$25~M$_\odot$. This picture of core-collapse supernova phenomenology has been proposed recently by \citet{Kuroda:2021eiv}. It also takes into account that the class of core-collapse supernova explosions of very massive progenitors, in the mass range of $M >30~M_\odot$, driven by the hadron-quark phase transition can also only be small fraction. Otherwise there would be a conflict with stellar population synthesis analyses, concerning the number of black holes in this mass range. Furthermore, the role of metallicity has yet incompletely been studied. From \citet{Fischer:2021} is becomes clear that the situation might be completely different at (extremely) low metallicity, usually connected with the early evolution of the universe.

\subsection{Observational signals of the hadron-quark phase transition}
One question remains, namely, is there an observable signature from the supernova explosions driven by the hadron-quark first-order phase transition? The present simulations based on multi-purpose hybrid EOS, applicable for core-collapse supernova studies, reveal that there are potentially observable signatures in both, the neutrino and the gravitational wave signals. These cannot only directly related with the formation of a quark matter phase. 

\subsubsection{Neutrino emission}
\begin{figure*}[t!]
\centering
\includegraphics[width=0.9\columnwidth]{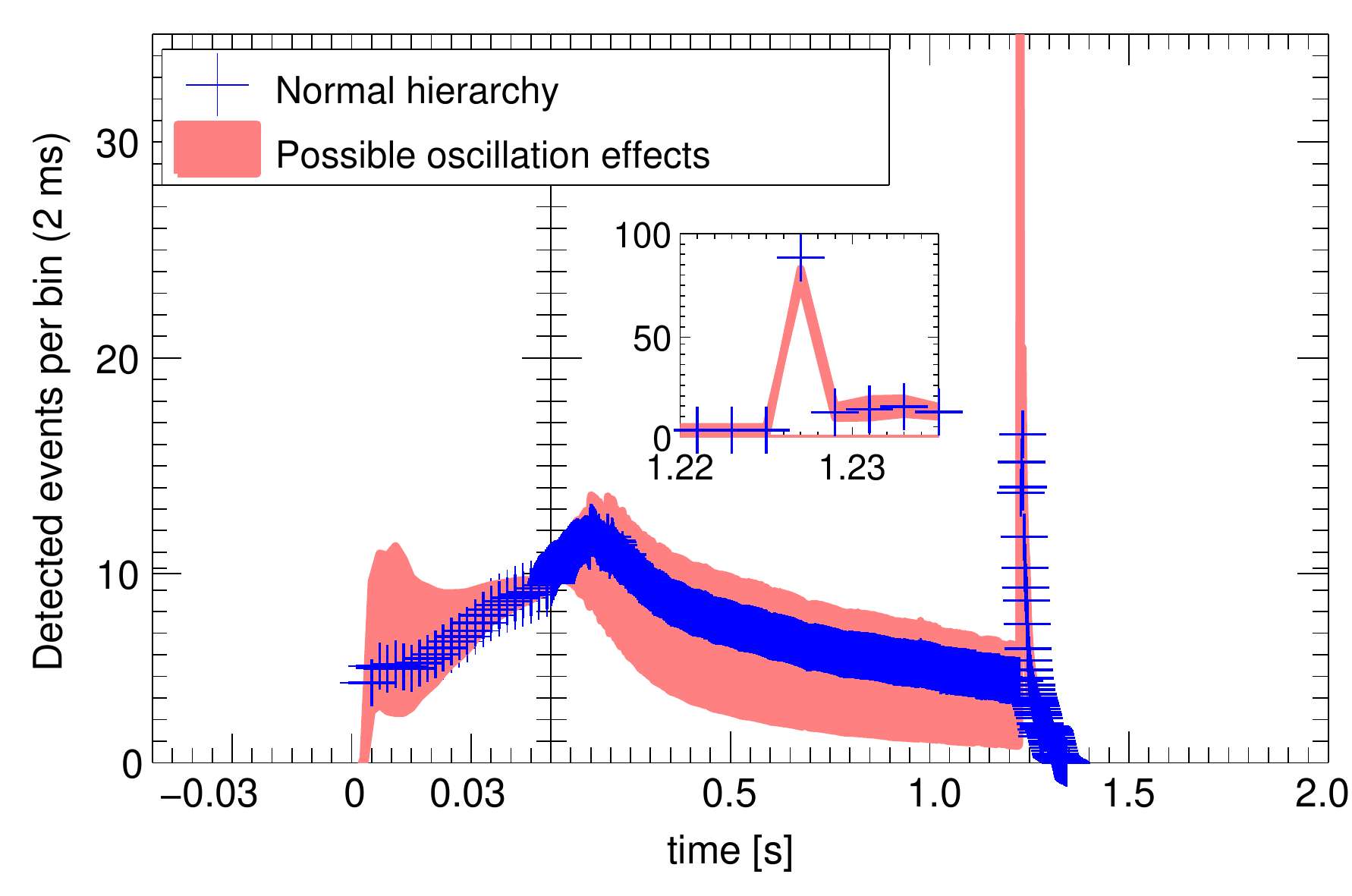}
\caption{\label{fig:sn_SK} \citep[Figure adopted from][]{Fischer:2018} Neutrino signal from a core-collapse supernova explosion launched from a 50~M$_\odot$ progenitor that takes into account the hadron-quark phase transition using the DD2F-RDF~1.2 hybrid EOS, evaluated at the Super-Kamiokande water Cherenkov detector at a fiducal distance of 10~kpc from the Earth.}
\end{figure*}

The emission of a burst-like signals is related with the dynamics of the formation and propagation of the second shock wave. This millisecond neutrino burst is released when the second shock wave, which forms a a direct consequence of the hadron-quark phase transition, propagates across the neutrino spheres of last inelastic scattering. This neutrino burst is detectable for a galactic event, as illustrated in Fig.~\ref{fig:sn_SK} examined at the Super-Kamiokande detector. This burst is absent in {\em standard} neutrino-driven core-collapse supernova explosions \citep[c.f.][]{Wu:2015}. Hence, if observed for a future galactic event, it represents a smoking gun signature for the appearance of quark matter. If not observed, one might be able to constrain the conditions, in terms of baryon density and temperature, up to which the hadron-quark phase transition can be excluded. Similar results have been obtained previously by \citet{Sagert:2008ka} and \citet{Fischer:2011}, however, employing a simplistic bag-model EOS for quark matter which is incompatible with the constrain deduced from the existence of massive pulsars of $\sim$2~M$_\odot$. 

\subsubsection{Emission of gravitational waves}
In general, core-collapse supernovae GWs stem from various different phases of the evolution. During the early post-bounce phase, high-amplitude GWs are emitted from convective motion, which lasts ups to several tens of milliseconds. After that, during the proceeding post-bounce mass accretion phase prior to the explosion onset, the GW emission is determined from a variety of aspects, e.g., neutrino-driven convection, the presence of hydrodynamics instabilities such as the standing accretion shock instability. One feature seen in all present core-collapse SN simulations is the continuously increasing GW frequency with time \citep[c.f.][and references therein]{Murphy09,BMuller13,KurodaT16ApJL,Vartanyan19b,Mezzacappa20,Shibagaki21}. This feature is considered to be associated with PNS oscillations as well as convection at the PNS interior at high density, which, in turn, excites the dominant post-bounce GW $f$-mode \citep[][]{Morozova18,Torres-Forne19,Sotani20b}. It is also present in the axially-symmetric SN simulations of \citet{OConnor:2020} and \citet{Kuroda:2021eiv} employing a hadron-quark hybrid EOS, however, prior to the phase transition when the post-bounce evolution is determined by the hadronic phase. 

\begin{figure*}[t!]
\centering
\includegraphics[width=0.75\columnwidth,angle=-90]{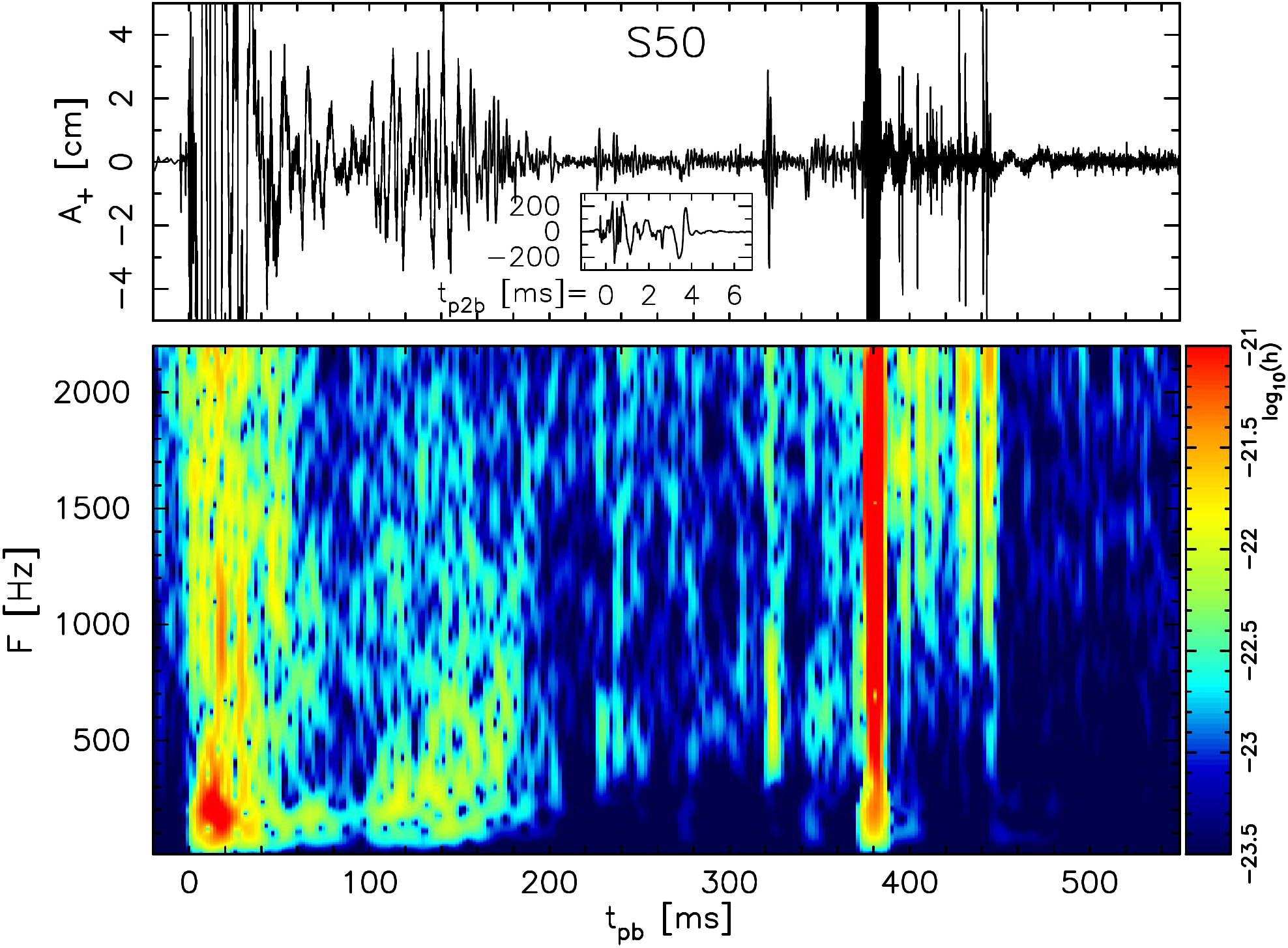}
\caption{\label{fig:sn_GW_spectra} \citep[Figure adopted from][]{Kuroda:2021eiv} 
Gravitational waveform $A_+$ (top panels) and spectrogram (bottom panels) for an axially-symmetric core-collapse SN simulation with hadron-quark hybrid EOS, launched from a 50~M$_\odot$ progenitor. The inlay at the top panel shows a zoom in of $A_+$ corresponding to the evolution shortly after the phase transition.
}
\end{figure*}

\begin{figure*}[t!]
\centering
\includegraphics[width=0.8\columnwidth,angle=-90]{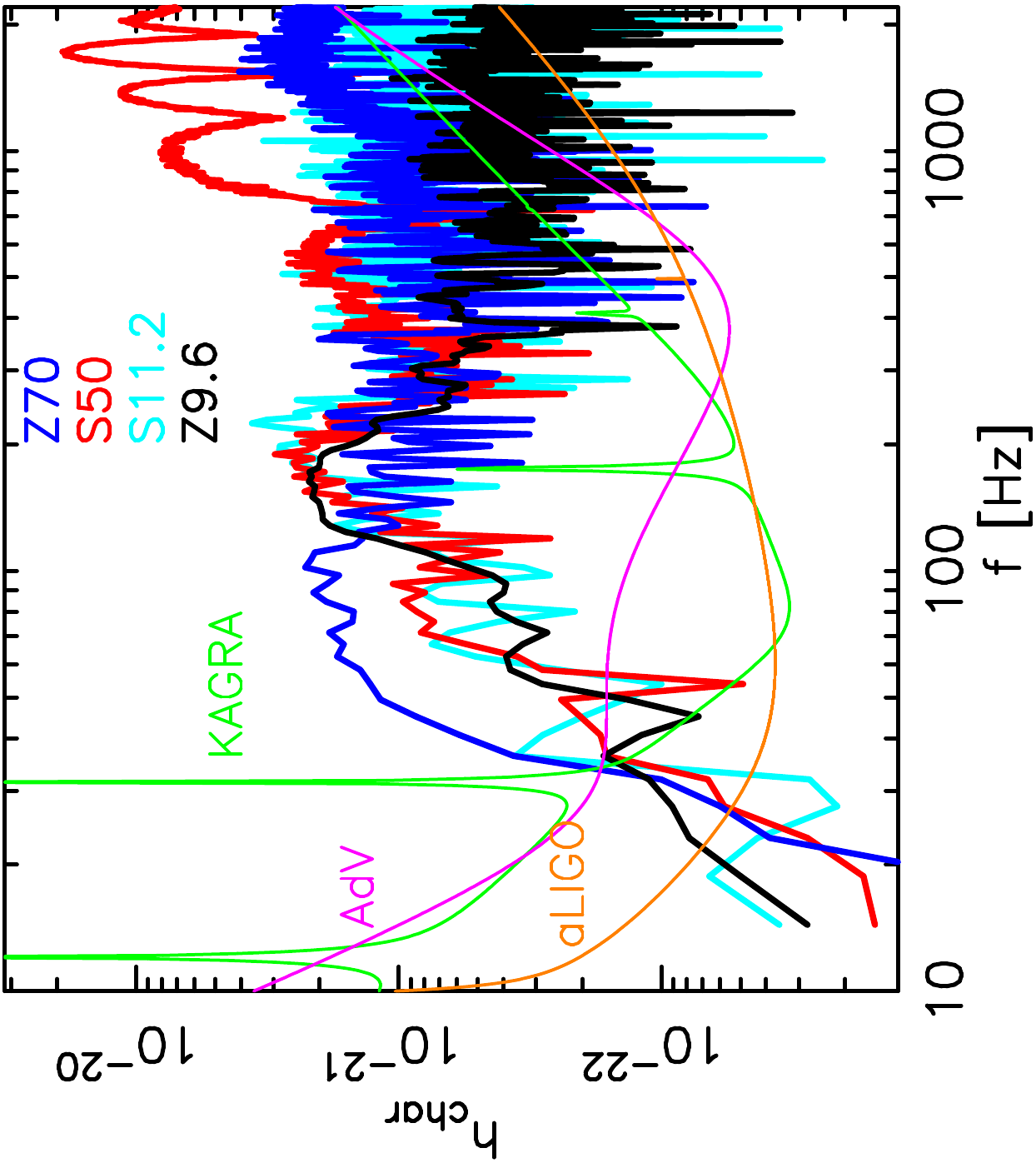}
\caption{\label{fig:sn_GW_detection} \citep[Figure adopted from][]{Kuroda:2021eiv} Gravitational wave signal, spectral amplitudes with respect to the frequency, for a sample of axially-symmetric core-collapse SN simulations launched from different progenitors: 9.6~M$_\odot$ (Z9.6), 11.2~M$_\odot$ (S11.2), 50~M$_\odot$ (S50) and 70~M$_\odot$ (Z70). All models are of solar metallicity except Z9.6 and Z70, which correspond to the class of extremely metal poor stars.}
\end{figure*}

Figure~\ref{fig:sn_GW_spectra} shows the post-bounce evolution of the GW amplitude (top panel) for an axially-symmetric SN simulation of \citet{Kuroda:2021eiv}, launched from a 50~M$_\odot$ progenitor employing the hadron-quark hybrid EOS of \citet{Fischer:2018}. Shown is only the non-vanishing component in axial symmetry observed along the equatorial plane, i.e., perpendicular to the symmetric axis. Note that since the SN simulations employ an effective treatment of General Relativity, the GW amplitude $A_+$, more precisely the GW strain, is obtained from the standard quadrupole formula \citep[for details, see][and references therein]{Shibata&Sekiguchi03,KurodaT14}. The bottom panel of Fig.~\ref{fig:sn_GW_spectra} shows the spectrogram of the GW  emission, which is obtained using a short time Fourier analysis. It shows the aforementioned continuously rising dominant $f$-mode frequency, starting at around 50--100~Hz rising to several hundreds of Hz. The sudden rise of the GW amplitude and frequency, at around 400~ms post bounce, is  the occurence of the hadron-to-quark matter phase transition and the associated PNS collapse, second bounce and formation of the second shock wave. It lasts for only few milliseconds. After that, the  PNS oscillations continue for several tens of milliseconds, which are responsible for the GW emission between about 400--450~ms, shown in Fig.~\ref{fig:sn_GW_spectra}. 

In order to address the question about the detectability of such GW signal, in particular the possibility to distinguish it from SN that are not associated with the hadron-quark phase transition, \citet{Kuroda:2021eiv} compared the GW emission with those of canonical, neutrino-driven SN explosions as well as one failed SN model in which case a black hole forms instead. The latter is launched from a low-metallicity 70~M$_\odot$ progenitor (Z70). In comparison to this one, as well as two other neutrino-driven SN models launched from a low-metallicity 9.6~M$_\odot$ (Z9.6) and a solar-metallicity 11.2~M$_\odot$ progenitor (S11.2), the high-frequency kHz GW signal is enhanced by nearly one order of magnitude for the 50~M$_\odot$ SN simulation with hadron-to-quark matter phase transition (see Fig.~\ref{fig:sn_GW_detection}). The origin of these high amplitudes is PNS oscillations and PNS convection after the phase transition took place. It is absent for Z70 because a black hole forms about 1~ms after the phase transition. The future observation of such a  high GW amplitude component at kHz frequencies, which is likely to be absent in all other SN, represents a smoking gun signature for the presence of a hadron-to-quark matter phase transition 
{\em and} a resulting stable remnant compact star with quark matter at the interior. This might be used as constraint for the presence of quark matter at high baryon densities, complementary to lattice QCD and experimental heavy-ion collision programs.

One caveat in predicting the GW emission from axially symmetric simulations might arise in the overestimation of convective motion due to the presence of a fixed symmetry axis. This aspect tends to overestimate the GW amplitude. Another, yet incompletely understood feature is the presence of a quiescent phase of GW emission during the post-bounce evolution (present between 200--350~ms post bounce in Fig.~\ref{fig:sn_GW_spectra} prior to the phase transition induced sudden rise of the amplitude). It has been speculated whether it corresponds to the period when the shock wave gradually recedes. Only with simulations based on three spatial dimensional setup will enable us to resolve this issue. 

\subsubsection{Nucleosynthesis yields of heavy elements}
The final phase of a supernova evolution after the supernova explosion onset, is characterized by the deleptonization of the nascent PNS, i.e. the diffusion of the trapped neutrinos. It lasts for several tens of seconds and is associated with the ejection of a low-mass outflow known as the neutrino-driven wind, which has long been studied as a possible site for the nucleosynthesis of heavy elements in the universe \citep[c.f.][and references therein]{Thielemann:2011PrPNP,Cowan2021}. It depends sensitively on the nucleosynthesis conditions of the neutrino-driven wind, i.e., entropy per particle and the neutron excess, as well as on their evolution \citep[][]{Woosley:1994ux}. Recent investigations have confirmed that canonical, neutrino-driven core-collapse SN explosions cannot yield a strong $r$-process nucleosynthesis component \citep[c.f.][]{MartinezPinedo:2014,Fischer:2020PhRvC}. 
The reason for this are the generally too low entropies per particle, on the order of 50--100~$k_{\rm B}$, and insufficient neutron excess or even generally proton-rich conditions if PNS convection is considered \citep[][]{Roberts:2012PhRvL,Mirizzi:2016}. No elements heavier than molybdenum, with atomic number $Z=42$, can be produced. The key input physics in the SN simulations is three-flavor Boltzmann neutrino transport, for the accurate prediction of the neutrino fluxes and spectra as well as their evolution, which determine nearly entirely these aforementioned nucleosynthesis conditions \citep[][]{Qian:1996xt}. Contrary, studies that are lacking neutrino transport have to assume the nucleosynthesis conditions. Furthermore, recent multi-dimensional SN simulations point to the presence of late-time mass accretion, i.e., after the successful SN explosion onset, which lasts for seconds \citep[c.f.][and references therein]{Mueller:2017,Bollig:2021,BurrowsVartanyan:2021Natur,NagakuraBurrowsVartanyan:2021}. The exception might be the class of neutrino-driven explosions of low-mass stellar progenitor in the mass of 8--9~M$_\odot$ \citep[][]{RadiceBurrows:2017,StockingerJanka:2020MNRAS} as well as those which originate from binary stellar evolution leading to stripped envelope SN explosions \citep[][]{MuellerTauris:2019}. If large down flows of matter persist after the SN explosion onset, then the neutrino-driven wind is likely to be absent and the nucleosynthesis results of these model might be very different from those of the neutrino-driven wind \cite{Witt:2021}. 

\begin{figure*}[t!]
\centering
\includegraphics[width=0.9\columnwidth,angle=0]{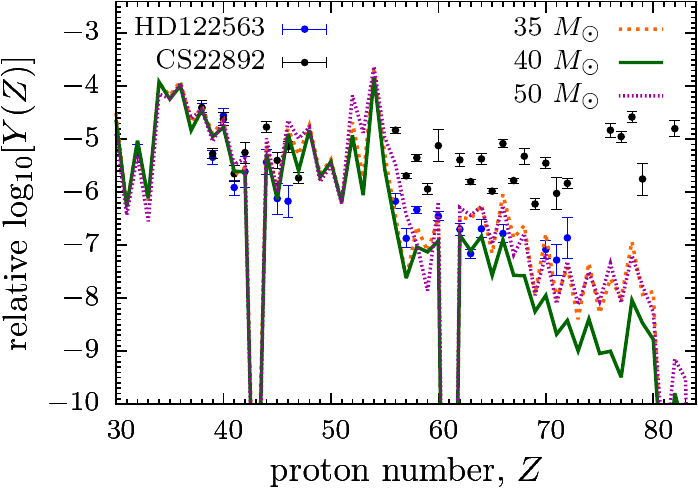}
\caption{\label{fig:sn_nucleosynthesis} \citep[Figure adopted from][]{Fischer:2020} Integrated elemental yields from three QCD driven core-collapse supernova explosions, launched from 35, 40 and 50~M$_\odot$ progenitors. For comparison, observations are shown for two metal-poor stars, HD122564 from \citet{Roederer:2012dr} and CS22892 from \citet{Sneden:2003zq}.}
\end{figure*}

The conditions found in SN explosions that are triggered through the hadron-to-quark matter phase transition by \citet{Fischer:2020} are very different from those of neutrino-driven SN explosions. One finds:
\begin{enumerate}
\item[ {\em (a)}] an early, low-entropy and neutron-rich component with electron fraction of $Y_e\simeq0.2-0.25$, which is ejected dynamically together with the expanding explosion shock wave,
\item[{\em (b)}] an intermediate component characterized by rising entropy and $Y_e$, and 
\item[{\em (c)}] a late-time component. 
\end{enumerate}
The latter is the high-entropy neutrino-driven wind, featuring an entropy per baryon on the order of several 100~$k_{\rm B}$, however, being subject of large uncertainties as pointed out before. The integrated nucleosynthesis results of three SN explosions driven by the QCD phase transition are shown in Fig.~~\ref{fig:sn_nucleosynthesis}, for three different progenitors of 35, 40 and 50~M$_\odot$, showing a robust strong $r$-process component with the production of elements not only corresponding to the third $r$-process peak at mass number $A\simeq195$ but even transactinides such as plutonium. However, the abundances of the rare-Earth elements as well as those of the third $r$-process peak are suppressed by several orders of magnitude in comparison to the second $r$-process peak at nuclear mass number $A\simeq 125$. Hence, this class of massive star explosions cannot account for the bulk part of $r$-process nucleosynthesis in the galaxy. Instead, it is as a rare $r$-process site, as was the considered in \citet{FarouqiThielemann:2021}, together with the class of magneto-rotationally driven core-collapse SN \citep[][]{LeBlanc:1970kg,Winteler:2012,Moesta:2014ApJL,Moesta:2015Natur}. The latter requires a high magnetic field strength of the progenitor in combination with rapid rotation, which is likely to be realized only for a subset of all core-collapse SN.

\newpage
\section{Quark matter in binary neutron star mergers}
\label{sec1.4}
For the astrophysics community the very first unambiguous detection of a neutron star coalescence in August 2017 was a seminal event as the observations revealed a 'proto-type' neutron star merger, which settled many questions that had been open or tentative before~\citep[][]{TheLIGOScientific:2017qsa}. Furthermore, the detection highlights the potential to learn about the presence of deconfined quark matter in neutron stars, i.e. at densities of a few time nuclear saturation density. Merging neutron stars lead to a number of observable phenomena, which in principle can carry in imprint of quark deconfinement. This includes foremost the gravitational-wave signal but also electromagnetic counterparts, where various processes generate emission in different parts of the spectrum~\citep[][]{Abbott2017multi}.

\begin{figure*}[h!]
  \includegraphics[width=0.495\columnwidth]{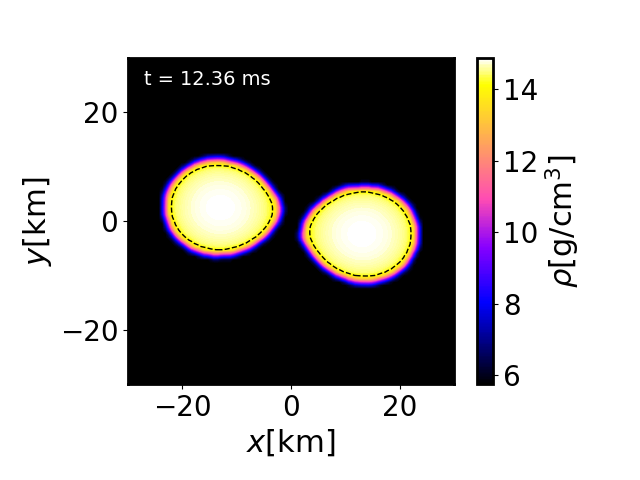}
  \includegraphics[width=0.495\columnwidth]{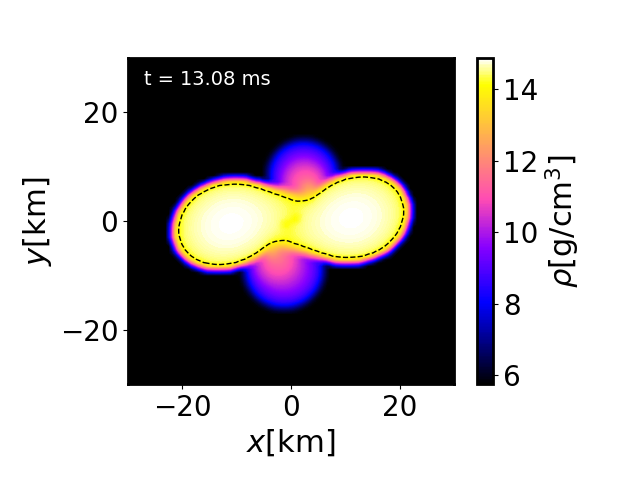}\\
  \includegraphics[width=0.495\columnwidth]{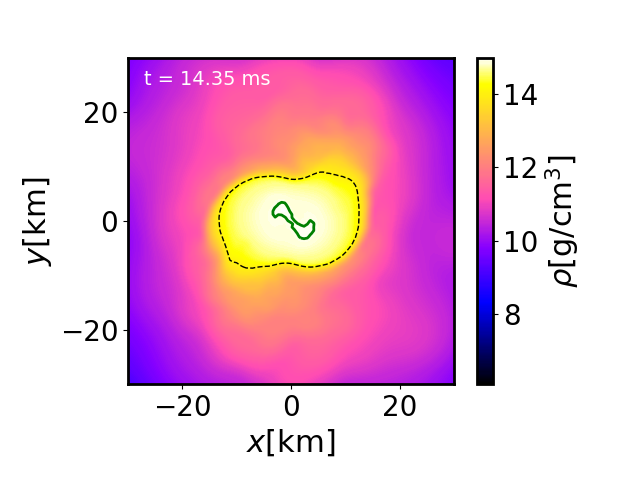}
  \includegraphics[width=0.495\columnwidth]{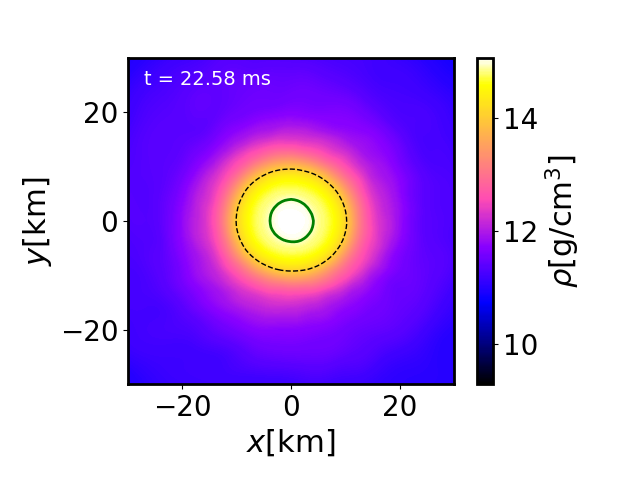}
  \caption{\label{fig:snap}  \citep[Figure from][]{Bauswein:2019skm} Density evolution of a neutron star merger with a hybrid equation of state with a phase transition to deconfined quark matter. Lines display density contours of  $\rho=10^{14}~\mathrm{g/cm^3}$ (dashed) and $8.7\times10^{14}\mathrm{g/cm^3}$ (solid). The latter density roughly corresponds to the onset density of the phase transition at zero temperature.}
\end{figure*}

The merger of two neutron stars proceeds through different stages, starting with a long-lasting ``inspiral'' phase, where the binary looses angular momentum and energy as a result of gravitational-wave emission, which leads to a decrease of the orbital separation. The inspiral continuously accelerates and finally leads to the coalescence of the binary components as a highly dynamical process with typical time scales of milliseconds (see Fig.~\ref{fig:snap}). For higher total binary masses the merger forms a black hole-torus system because the merger remnant cannot be stabilized against the gravitational collapse. Recall that typical neutron star masses are about 1.3 to 1.4~$M_\odot$ such that the mass of the merged object will usually exceed the maximum mass of nonrotating neutron stars. If the total binary mass is below a certain threshold mass, the merging binary components create a neutron star merger remnant. Since most neutron star binaries may have a total mass of about 2.7~$M_\odot$ and since the threshold mass for black hole formation is likely in the range between 2.8 and 3.4~$M_\odot$, one may expect that the typical outcome of a merger is the formation of a neutron star remnant. This object is rapidly rotating and initially highly deformed and strongly oscillates reaching a quasi-equilibrium on time scales of several 10 milliseconds to enter a phase of secular evolution.

During the merging process the densities increase as a more massive and more compact object forms. Therefore, the prospects for the occurrence of deconfined quark matter in the ``postmerger'' stage are generally better compared to the inspiral. Moreover, the temperature in the merger remnant increases and can reach several 10~MeV and in the shock-heated contact interface even about 100~MeV. This further enhances the likelihood for the hadron-quark phase transition to take place. Whether deconfined quark matter is present in the postmerger remnant or even before merging during the inspiral, is currently open and depends on the unknown temperature-dependent onset density of the hadron-quark phase transition. This motivates to identify specific signatures of quark deconfimenent in neutron star merger observables.

\subsection{Gravitational waves}
The gravitational wave signal of a neutron star merger reflects the stages of the coalescence, i.e.\ the inspiral and the postmerger evolution, and encodes the dynamics of the system, which are dominantly determined by the binary masses and the equation of state. In turn, the detailed analysis of the gravitational wave signal reveals the binary masses and information on the equation of state~\citep[see][]{Abbott2019}. The impact of the binary masses is described by two parameters. The chirp mass, defined as follows,
\begin{equation}
\mathcal{M}_\mathrm{chirp}=(M_1 M_2)^{3/5}/(M_1+M_2)^{1/5}~,
\end{equation}
with $M_1$ and $M_2$ being the masses of the individual binary components, has the strongest influence on the inspiral gravitational wave emission and is thus measured with high precision. The binary mass ratio $q=M_1/M_2$ affects the signal only during the last orbits before merging and can thus be inferred only with less precision. Combining $\mathcal{M}_\mathrm{chirp}$ and $q$ yields the physical masses of the system. 

Equation-of-state effects during the inspiral are described by the so-called combined tidal deformability, which is defined as follows \citep[c.f.][]{Flanagan2008,Hinderer2008},
\begin{equation}
\tilde{\Lambda}=\frac{16}{13}\frac{(M_1+12 M_2) M_1^4\Lambda_1  + (M_2+12 M_1) M_2^4\Lambda_2 }{(M_1+M_2)^5}~.
\end{equation}
Here $\Lambda_1$ and $\Lambda_2$ are the tidal deformabilities of the individual stars given by $\Lambda=\frac{2}{3}k_2\left(\frac{c^2\,R}{G\,M}\right)^5$ with the stellar radius $R$ and the tidal Love number $k_2$. $k_2$ is a stellar structure parameter, which is as the radius uniquely determined by the equation of state. A larger $\tilde{\Lambda}$ accelerates the inspiral since systems with a finite tidal deformability lead to stronger gravitational-wave emission compared to point-particle binaries of the same masses. Its definition illustrates that $\Lambda$ scales tightly with the stellar radius, and the constraint which GW170817 put on the tidal deformability excludes neutron stars in the given mass range with radii larger than about 13.5~km~\citep[][]{TheLIGOScientific:2018}. 

As the stellar radius, the tidal deformability is similarly affected by the presence of a phase transition to deconfined quark matter. Specifically, the hadron-quark phase transition can lead to a kink in the relation between tidal deformability and mass. The kink occurs at the neutron star mass $M_\mathrm{onset}$, where quark matter starts to appear. Hence, the presence of quark matter may be identified by detecting systems with different masses and measuring the respective tidal deformability~\citep[detailed can be found in][and references therein]{Chatziioannou2020,Chen2020,Pang2020PhRvR...2c3514P}. However, even for models which predict a rather strong impact of the phase transition on the stellar structure, the kink in $\Lambda(M)$ may not be very pronounced \citep[c.f.][]{Han2019PhRvD..99h3014H,Bauswein:2019skm,Christian2019PhRvD..99b3009C}. Therefore, identifying quark matter in neutron stars during the gravitational-wave inspiral would require very precise measurements of the tidal deformability, which seems challenging in regard to statistical and systematic uncertainties in the gravitational-wave data analysis.

The definition of the tidal deformability also shows that $\Lambda$ strongly decreases with the stellar mass since more massive stars become more compact and thus more similar to point particles. Hence, it is more challenging to measure finite-size effects in high-mass binaries. This implies that the identification of quark matter through the inspiral may be more intricate in particular if the onset density of the hadron-quark phase transition is relatively high. Although the currently very low number of neutron star merger detections does not yield good statistics on the population properties of neutron star binaries, one may expects that there will be fewer systems with high total binary masses. 

\begin{figure}[h!]
\centering
  \includegraphics[width=0.975\columnwidth]{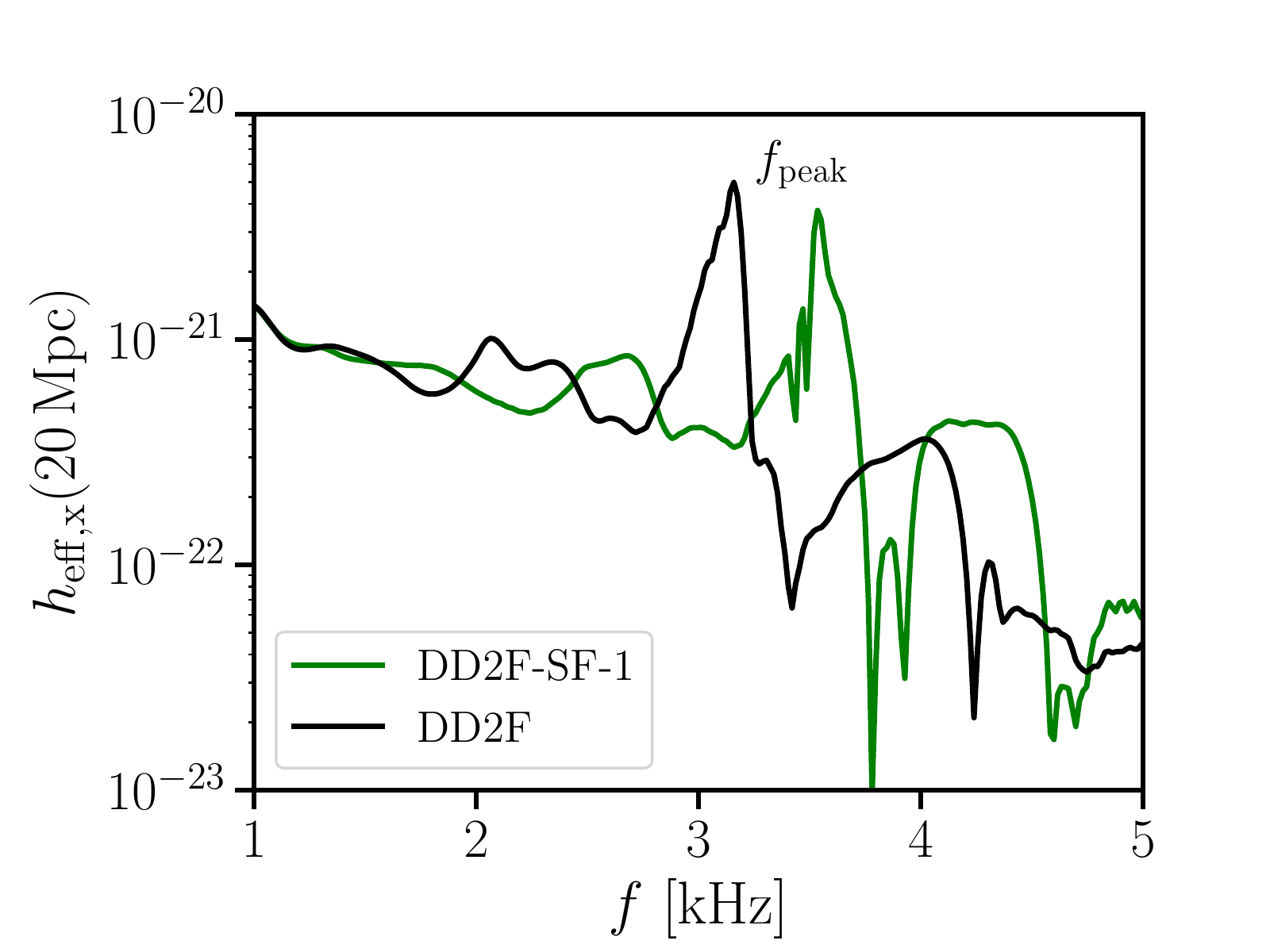}
  \caption{\label{fig:spectrum} \citep[Figure taken from][]{Bauswein:2019} Gravitational-wave spectrum of two neutron star merger simulations with a total mass of 2.7~$M_\odot$. The black curve adopts a purely hadronic equation of state, while the green curve displays the spectrum for a model undergoing a phase transition to deconfined quark matter.}
\end{figure}

In this regard the postmerger phase may be more promising since the remnant probes a density regime which is higher than that of the initial stars, and, thus, low-mass binaries may already reveal signals of quark matter. The equation of state affects the frequencies of the postmerger oscillations and therefore the frequencies of the postmerger gravitational-wave emission. Typical gravitational-wave spectra are shown in Fig.~\ref{fig:spectrum}, where one can clearly identify different peaks. The most prominent feature is a pronounced peak in the kHz range, which corresponds to the dominant oscillation mode of the remnant and is generated by the fundamental quadrupolar fluid mode. The frequency of this oscillation is determined by the compactness of the remnant, where more compact stellar objects oscillate at higher frequencies. This implies a the strong sensitivity of the frequencies on the equation of state since the properties of the equation of state shape the stellar structure of the remnant.

Figure~\ref{fig:spectrum} shows simulation results for an equation of state model with phase transition to quark matter (green) and one model without phase transition (black) for a merger of two stars with 1.35~$M_\odot$. Both models are based on the same hadronic equation of state at lower densities and only differ at higher densities, where the model with phase transition features a significant softening. This softening leads to a more compact remnant and thus to a higher postmerger oscillation frequency, while the general morphology of the spectra is similar. In the particular case displayed in Fig.~\ref{fig:spectrum} the phase transition takes place right after merging (see Fig.~\ref{fig:snap}) such that the inspiral proceeds identically.

\begin{figure}[h!]
\centering
  \includegraphics[width=0.975\columnwidth]{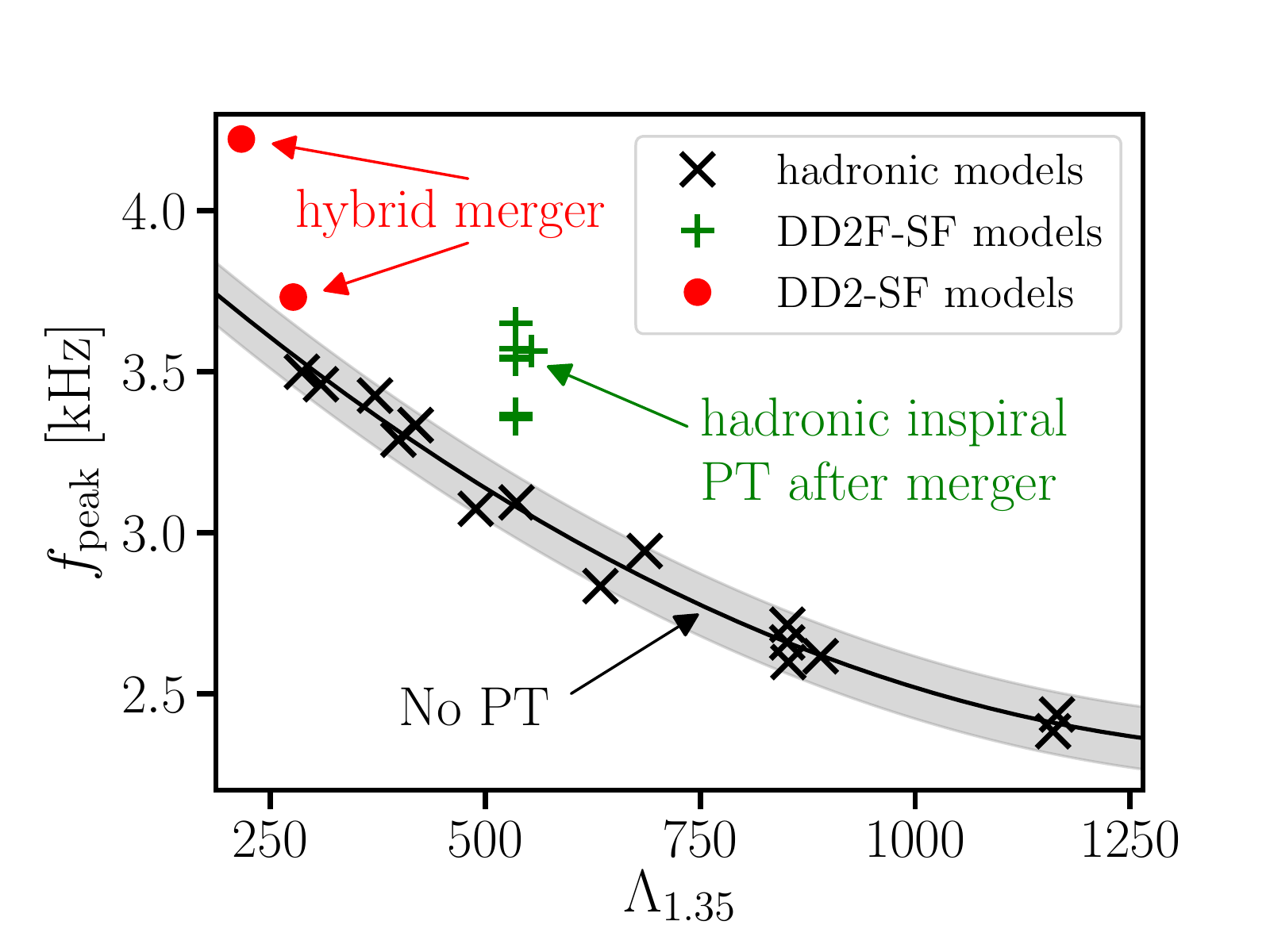}
  \caption{\label{fig:fpeaklam} \citep[Figure from][]{Bauswein2020} Postmerger gravitational-wave frequency as function of the tidal deformability for 1.35-1.35~$M_\odot$ binaries with many equation of state models. Purely hadronic models (crosses) follow a tight relation fitted by the black curve. Models undergoing a phase transition before merging (dots) or after merging (plus signs) lead to a frequency shift relative to the models without phase transition.}
\end{figure}

While a frequency shift by a phase transition is intuitive, it is not obvious that a high postmerger frequency alone is indicative of the presence of deconfined quark matter because also purely hadronic equation of state models can reach frequencies in the range up to 4~kHz for this binary configuration. However, an increased postmerger frequency is in fact characteristic of a phase transition if one compares $f_\mathrm{peak}$ to the tidal deformability inferred from the inspiral phase \citep[details can be found in][and references therein]{Bauswein:2019}. This comparison is displayed in Fig.~\ref{fig:fpeaklam}, which compiles simulation results for many calculations with  the same binary mass configuration but with different equation of state models with and without phase transition to deconfined quark matter.

All purely hadronic models follow a tight correlation between $f_\mathrm{peak}$ and $\Lambda_{1.35}$ (note that for equal-mass binaries $\tilde{\Lambda}=\Lambda_1=\Lambda_2$). This scaling is not fully unexpected because $\Lambda_{1.35}$ is a good proxy for the stellar structure of the inspiralling stars and consequently also for the postmerger remnant. Although densities increase during merging, purely hadronic equations of state do not change very abruptly such that $\Lambda_{1.35}$ does also describe the stellar structure of objects with densities exceeding those in 1.35~$M_\odot$ neutron stars.

In contrast, the occurrence of deconfined quark matter can change the equation of state strongly beyond the onset density of the phase transition. Consequently, the softening of the equation of state and the resulting frequency shift of the postmerger gravitational-wave emission is not captured by stellar parameters which are only sensitive to the density regime below the onset density (such as $\Lambda_{1.35}$ in Fig.~\ref{fig:fpeaklam}). This explains the characteristic offset of equation of state models with quark matter in Fig.~\ref{fig:fpeaklam}. Hence, a significant deviation of the postmerger gravitational wave frequency $f_\mathrm{peak}$ from the empirical relation $f_\mathrm{peak}(\Lambda_{1.35})$ for purely hadronic models (black line) does provide very strong evidence for the occurrence of deconfined quark matter in neutron stars \citep[see][]{Bauswein:2019}.

Some of the purely hadronic models in Fig.~\ref{fig:fpeaklam} feature a phase transition to hyperonic matter, which apparently does not soften the equation of state sufficiently to yield a strong frequency deviation. Hence, a characteristic frequency shift relative to $\Lambda_{1.35}$ does indeed represent an unambiguous signature of quark matter.

It is conceivable that the frequency shift is a signature which also occurs for other hadronic equations of state describing the density regime below the phase transition \citep[see][and references therein]{Bauswein:2019,Liebling2020,Blacker2020,Weih2020,Hanauske2021EPJST.230..543H,Prakash:2021wpz}. The models depicted by plus signs in Fig.~\ref{fig:fpeaklam} are based on the same hadronic equation of state but make different assumptions about the properties of the hadron-quark phase transition and the properties of deconfined quark matter in the non-perturbative regime beyond the phase transition. Therefore, the models occur at the same tidal deformability $\Lambda_{1.35}$ but result in differently strong frequency shifts\footnote{One model with a slightly higher $\Lambda_{1.35}$ features a small modification of the purely hadronic equation of state.}. In fact, the magnitude of the frequency increase encodes the ``strength'' of the phase transition. For instance, for a roughly similar onset density and stiffness of the quark matter phase, the frequency shift scales with the density jump across the phase transition \citep[see][]{Bauswein:2019}. Generally, it is clear that the signature of quark matter will be more pronounced for models which feature a stronger deviation from the corresponding hadronic reference model. This may also imply that for certain hybrid models the aforementioned signature of a relative frequency increase may not be very pronounced or even absent if the equation of state closely resembles the behavior of a purely hadronic model. See also the model in~\citet{Most:2019PhRvL}, where quark matter exhibits only a very weak impact on the gravitational-wave frequency but influences the collapse time scale of the remnant.

Moreover, the frequency shift may not be very strong if only a small quark matter core forms in the center of the remnant for binary configurations which are barely massive enough to reach conditions in the postmerger remnant to undergo a phase transition~\citep[see][]{Blacker2020}. In this case, however, a more massive binary with a sizeable quark matter core may yield a very strong signature. \citet{Blacker2020} provides an extensive study of the impact of quark matter in neutron star mergers for different binary mass configuration including unequal-mass systems, which can similarly yield characteristic frequency shifts. Depending on the specific properties of the phase transition and the quark matter equation of state, the occurrence of deconfined quark matter during the postmerger evolution may be delayed, as was discused in \citet{Weih2020}.

The dots in Fig.~\ref{fig:fpeaklam} demonstrate that a frequency increase relative to the empirical $f_\mathrm{peak}(\Lambda_{1.35})$ relation may also occur for mergers of hybrid stars, i.e. systems which contain quark matter cores already before the coalescence during the inspiral~\citep[][]{Bauswein2020}. 

Figure~\ref{fig:fpeaklam} illustrates that a relatively coarse determination of the tidal deformability of about $\Delta\Lambda\approx 100$ may be sufficient to identify the occurrence of the hadron-quark phase transition through a postmerger frequency shift. The dominant postmerger gravitational-wave frequency was not detected in GW170817. However, simulations show that relatively minor upgrades of the sensitivity of existing gravitational-wave instruments in the kHz range are sufficient to detect the postmerger peak of a signal similar to GW170817~\citep[details can be found in][]{Chatziioannou2017,Torres2019,Wijngaarden2022}. These gravitational-wave data analysis studies reveal that $f_\mathrm{peak}$ can be measured within a few 10~Hz, which is sufficiently precise to detect sizeable deviations from the $f_\mathrm{peak}(\Lambda)$ relation for purely hadronic models.

These considerations imply that any sufficiently precise measurement of $f_\mathrm{peak}$ and the tidal deformability will be informative about properties of the hadron-quark phase transition. While a significant postmerger frequency shift will provide very strong evidence for the presence of deconfined quark matter, data compatible with the $f_\mathrm{peak}(\Lambda)$ relation for purely hadronic models rules out the presence of a strong phase transition in the density regime of the merger remnant. This would thus place a constraint on the onset of quark deconfinement assuming that the presence of quark matter would lead to a {\it strong} signature as described above.

The probed density regime by the postmerger remnant and thus the concrete constraint that can be put on the onset density of the hadron-quark phase transition, is also accessible by gravitational wave parameters. For purely hadronic equation of state models there exist tight empirical relations between the tidal deformability or the postmerger gravitational wave frequency $f_\mathrm{peak}$, respectively, and the maximum density in the merger remnant~\citep[see][]{Bauswein:2019,Blacker2020}. Those universal relations hold for a large range of binary masses and can be employed to infer an upper or lower limit on the onset density of the hadron-quark phase transition depending on whether the comparison between $f_\mathrm{peak}$ and $\Lambda$ for the given binary masses reveals evidence for the presence or absence of quark matter.

Two complications may arise, which have been stressed in detail in~\citet{Blacker2020}. If the quark matter core is too small, the impact on the stellar structure and thus postmerger gravitational wave emission will not be very pronounced but only occur for more massive binary systems. Moreover, the presence of quark matter may not in any case lead to a sizeable effect if for instance the properties of the quark matter equation of state are mimicking those of hadronic matter as in~\citet{Alford2005}\footnote{If values of $f_\mathrm{peak}$ and $\Lambda$ are compatible with $f_\mathrm{peak}(\Lambda)$ of hadronic matter, either no phase transition occurred or quark matter properties are very similar to hadronic matter. Hence, the sketched procedures can only rule out the presence of a {\it strong} phase transition up to a given density and thus constrain the onset density of a strong phase transition~\citep[][]{Blacker2020}.}. The latter issue requires further investigations to reveal possible other signatures of quark matter that can inform about the presence of quark matter in the case its impact on the stellar structure is relatively weak.

\subsection{Mass ejection and electromagnetic counterparts}

Gravitational waves are not the only messenger from neutron star mergers. Apart from gamma-ray bursts and associated electromagnetic emission in different parts of the spectrum, mass ejection during the merger process can lead to electromagnetic radiation in the infrared, optical and ultraviolet. Material ejected from a neutron star merger expands with high velocities (up to a few 10 per cent of the speed of light) and undergoes the rapid neutron capture process forming heavy elements~\cite{Cowan2021}. The freshly synthesized elements are radioactive and the subsequent decays deposit heat in the initially opaque expanding ejecta cloud. The thermal emission of this matter produces an electromagnetic counterpart commonly called kilonova, which evolves on time scales of hours and days and becomes bright enough to be observable with telescopes~\cite{Metzger2019}. In the case of GW170817 the gravitational wave signal provided an estimate of the sky localization, and the corresponding kilonova was found several hours after the initial alert circulated after the detection by the gravitational wave instruments~\citep[][]{Abbott2017multi}.

The properties of the electromagnetic emission yield a coarse estimate of the ejecta parameters like their mass and outflow velocities. The brightness of the kilonova for instance scales approximately with the amount of ejecta, although the quantitative interpretation is very challenging because of various uncertainties, e.g. about atomic data of heavy nuclei, and different ejection mechanisms playing a role on different time scales and leading to different ejecta components. The ejecta characteristics depend on the merger dynamics and therefore on the binary masses and the equation of state~\citep[c.f.][and references therein]{Hotokezaka2013,Bauswein2013a}. Hence, kilonova properties in principle contain information on the equation of state and thus possibly on the presence of quark matter in neutron stars. 

\begin{figure}[h!]
\centering
  \includegraphics[width=0.975\columnwidth]{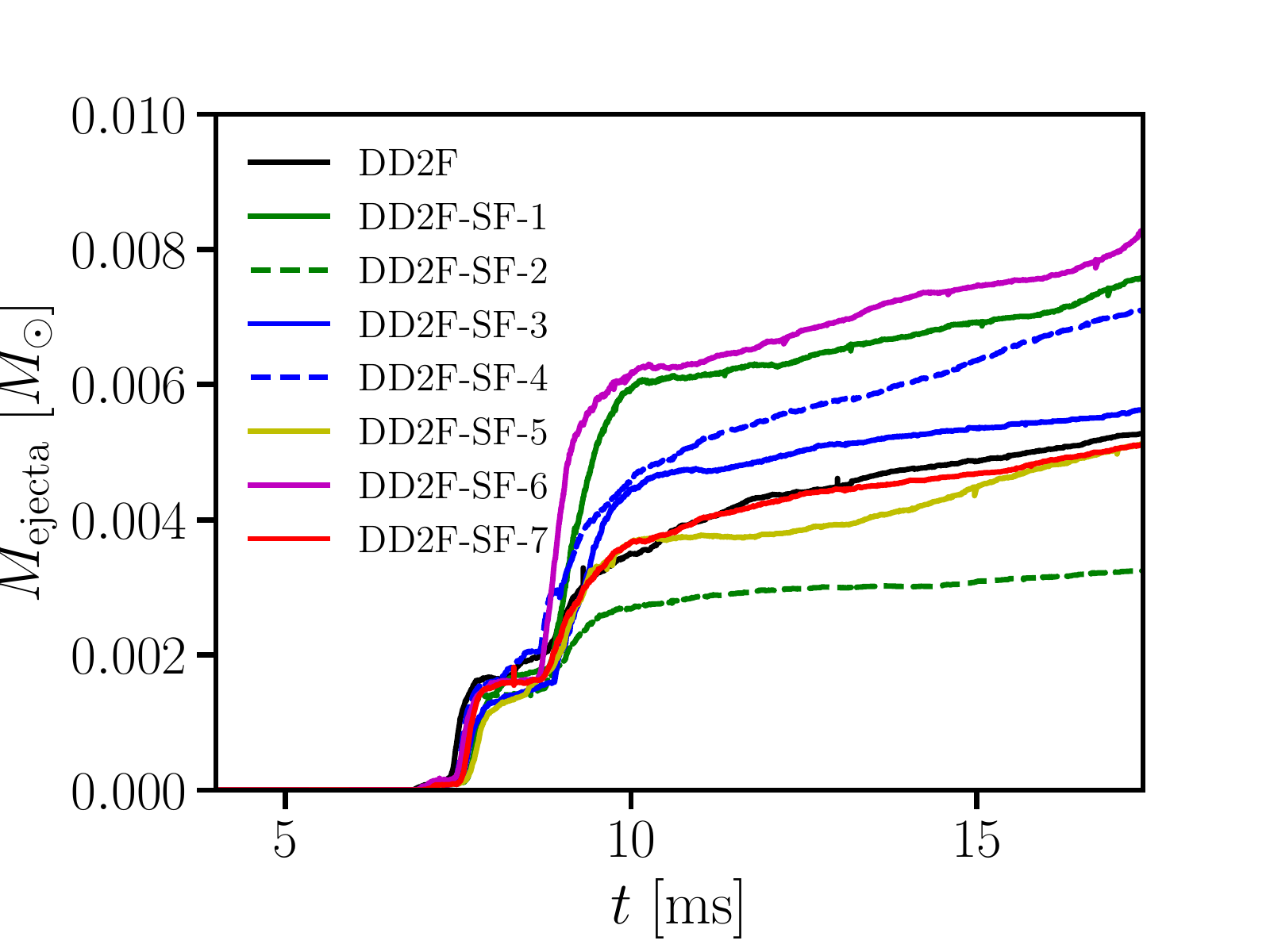}
  \caption{\label{fig:mej} \citep[Figure taken from][]{Bauswein:2019skm} Amount of unbound matter as function of time in simulations of 1.35-1.35~$M_\odot$ binaries for different hybrid equations of state. The black curve shows the purely hadronic reference model without phase transition. }
\end{figure}

Figure~\ref{fig:mej} displays the amount of unbound matter as function of time for 1.35-1.35~$M_\odot$ mergers. The calculations run only for a few 10~ms after merging and thus do only provide the initial dynamical ejecta component, while matter becoming unbound during the subsequent secular evolution is not captured. The black curve shows the result for a purely hadronic equation of state model, while all other curves exhibit the ejecta mass for hybrid equations of state undergoing a phase transition to deconfined quark matter. This set of models demonstrates that there is no very distinct and unique impact of the hadron-quark phase transition on mass ejection in neutron star mergers, as some models lead to enhanced ejecta masses while others yield a comparable or even reduced amount of unbound material in comparison to the purely hadronic reference equation of state (black line).

The scatter among the different models visible in Fig.~\ref{fig:mej} is to be compared to ejecta mass variations of various purely hadronic equations of state. Neutron star merger simulations for a larger sample of purely hadronic equation of states show roughly comparable variations of the ejecta mass even for models which result in similar stellar parameters~\citep[see][]{Bauswein2013a}. This implies that it is very difficult to unambiguously deduce the occurrence of quark matter from kilonova observations alone since bulk ejecta properties apparently do not carry unambiguous signatures of quark matter~\cite{Bauswein:2019skm,Prakash:2021wpz}. Moreover, nuclear network calculations postprocessing merger simulations do not reveal significant differences in the resulting elemental abundance patterns comparing models with and without phase transition~\citep[][]{Prakash:2021wpz}, although much more work is required to properly address the full complexity of this point.

\subsection{Black-hole formation}
Albeit kilonova observations may not be immediately informative about the occurrence of deconfined quark matter, they do contain information about the merger outcome. All models displayed in Fig.~\ref{fig:mej} yield a neutron star merger remnant since their total binary mass of 2.7~$M_\odot$ is below the respective threshold mass $M_\mathrm{thres}$ for prompt black-hole formation. If the total binary mass exceed the threshold for prompt black-hole formation, the ejecta mass and thus the kilonova brightness is strongly reduced~\cite{Bauswein2013a}. Hence, combining the binary masses measured by the gravitational wave inspiral and the interpretation of the kilonova emission can yield an estimate of the threshold mass $M_\mathrm{thres}$ for direct black hole-formation. Along these lines, it has been argued that GW170817 did not result in a prompt collapse. Also, the presence or absence of postmerger gravitational-wave emission informs about a gravitational collapse after merging. In essence, it is conceivable that $M_\mathrm{thres}$ can be determined or at least constrained from future observations.

This poses the question whether the occurrence of deconfined quark matter affects the threshold mass $M_\mathrm{thres}$ for black-hole formation. It is intuitively clear that the stability of a neutron star system in general and of $M_\mathrm{thres}$ in particular will strongly depend on the properties of the equation of state. Numerically $M_\mathrm{thres}$ can be determined by performing several simulations for different binary masses for a given equation of state and by checking the respective outcome. Such estimates for a large sample of different purely hadronic equation of states yield $M_\mathrm{thres}$ in a relatively broad range between roughly 2.8 and 3.8~$M_\odot$ depending on the specific properties of the microphysical model, as was discussed in \citet{Bauswein2020PhRvL.125n1103B}.

\begin{figure}[h!]
\centering
  \includegraphics[width=0.975\columnwidth]{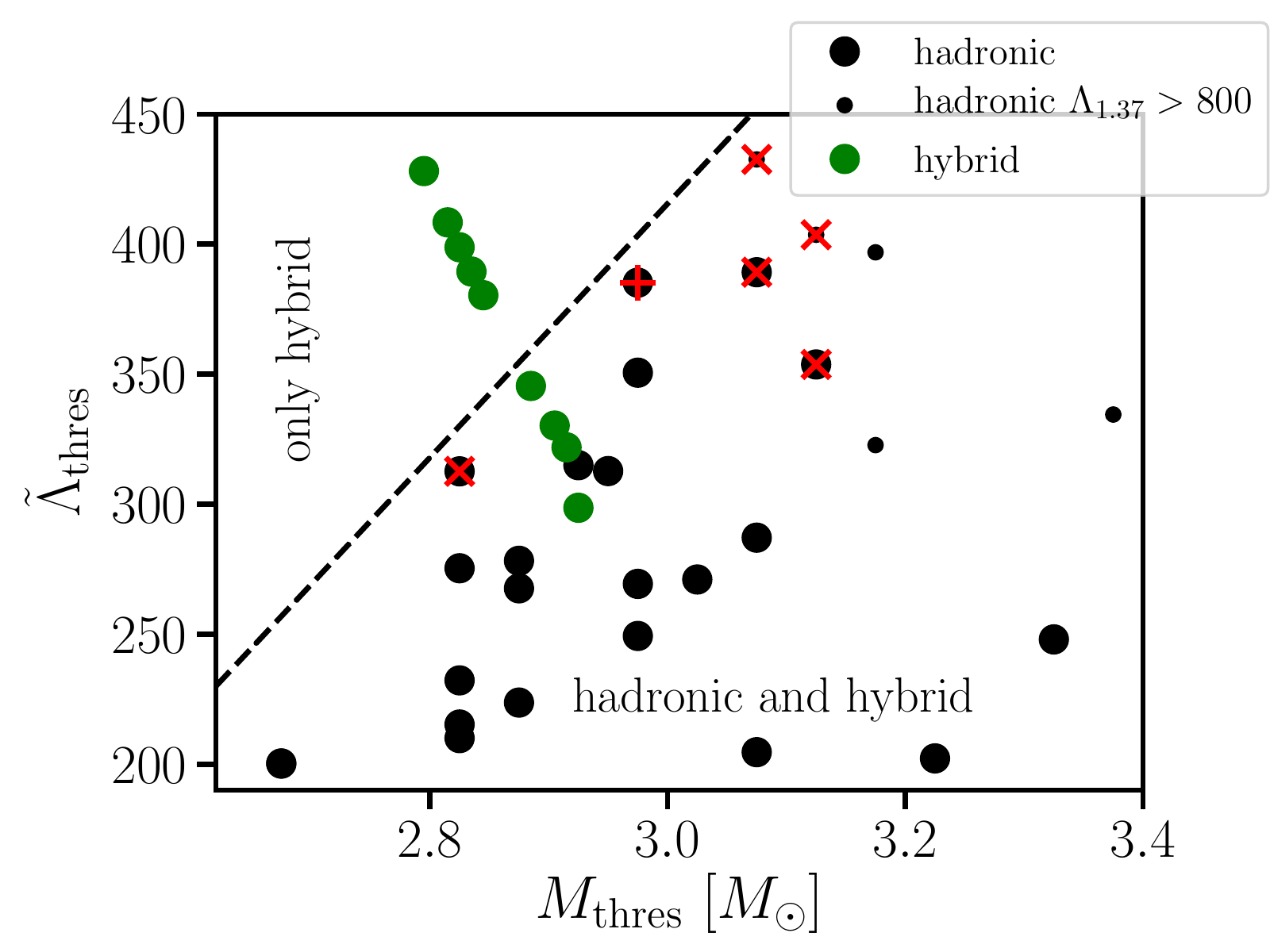}
  \caption{\label{fig:collapse} Collapse behavior of purely hadronic models (black) and hybrid equations of state with phase transition to deconfined quark matter (green). $M_\mathrm{thres}$ is the threshold mass for prompt black-hole formation. $\Lambda_\mathrm{thres}$ is the tidal deformability of the binary with a total mass of $M_\mathrm{thres}$, i.e. the tidal deformability of the system with a marginally stable remnant for the given equation of state. Purely hadronic equations of state lead to threshold properties below the black dashed line.}
\end{figure}

A $M_\mathrm{thres}$ estimate may yield signatures of quark matter in neutron stars since the occurrence of a phase transition can but does not necessarily need to result in a destabilization of the merger remnant~\cite{Bauswein2020PhRvL.125n1103B}. This is summarized in Fig.~\ref{fig:collapse}, which displays $M_\mathrm{thres}$ versus $\Lambda_\mathrm{thres}$ for different equations of state. $\Lambda_\mathrm{thres}$ is the tidal deformability measured in a system with a total mass of $M_\mathrm{tot}=M_\mathrm{thres}$ and can be inferred from the gravitational-wave inspiral signal. Hence, future observations will narrow down the area where the true equation of state lies in Fig.~\ref{fig:collapse}.

It is striking that in Fig.~\ref{fig:collapse} purely hadronic models are located below the dashed line, whereas some hybrid equations of state with phase transition to deconfined quark matter occur in the upper left corner of the diagram. Hence, some hybrid models feature a relatively low  $M_\mathrm{thres}$ in comparison to the tidal deformability. This may not be surprising because in these cases the tidal deformability is determined by the hadronic regime of the equation of state below the onset of the phase transition, whereas the appearance of quark matter during the merging softens the equation of state and destabilizes the remnant, i.e. reduces $M_\mathrm{thres}$. The softening of the equation of state and the associated reduction of $M_\mathrm{thres}$ is not captured by the tidal deformability since it is only sensitive to the equation of state at lower densities. This effect is somewhat similar to the shift of the postmerger gravitational-wave frequency as result of a strong compactification of the remnant. Such sudden changes of the equation of state do not occur for purely hadronic models, which is why they cannot occur much beyond the dashed line in Fig.~\ref{fig:collapse}.

Identifying the presence of quark matter through the collapse behavior in Fig.~\ref{fig:collapse} requires only a moderate precision of a tidal deformability measurement in comparison to resolving directly the kink induced by a phase transition in $\Lambda(M)$. Moreover, a single measurement may already provide evidence for quark matter if $\Lambda_\mathrm{thres}$ and $M_\mathrm{thres}$ are correspondingly constrained. As apparent from Fig.~\ref{fig:collapse} the comparison between $\Lambda_\mathrm{thres}$ and $M_\mathrm{thres}$ may not in any case reveal the presence of quark matter as some hybrid models can mimic the behavior of purely hadronic equations of state.

\newpage
\section{Summary} 
\label{sec1.5}
In this contribution, we have illustrated 
that there exist pronounced and unambiguous features which reveal the occurrence
of deconfined quark matter in 
supernova explosions and in
neutron star mergers. 
\\
In supernova simulations, the quark matter deconfinement phase transition leads to an explosion mechanism for massive blue supergiant stars which without the phase transition would evolve into black holes and represent failed supernovae.
We have demonstrated that as a result of the deconfinement-driven explosion mechanism neutron stars with high masses $\gtrsim 2~M_\odot$ are born, while the population of this mass segment within the standard neutrino-driven supernova explosion is quite unlikely, see \citet{Muller:2016ujh}.
As observational signatures we have discussed a second neutrino signal which appears about 1.2 ms after bounce and would be detectable by Super-Kamiokande if the supernova goes off in our galaxy,
as well as a gravitational wave signal in the spectrogram of an axially-symmetric core-collapse supernova about 400 ms post bounce.
It has also been found that a supernova exploding by the quark deconfinement mechanism must be a relatively rare event because it cannot account for the bulk part of the r-process nucleosynthesis in the galaxy due to a strong suppression of the synthesis of rare-earth elements and those of the third r-process peak. 
\\
Most of the signals of quark deconfinement in binary neutron star mergers are connected to the postmerger evolution, which may not be unexpected since the merger remnant probes higher densities than the inspiralling stars and thus naturally carries a stronger imprint of the equation of state at higher densities. 
We have focussed on the discussion of two deconfinement signals:
the shift of the peak frequency of postmerger gravitational wave signal (occurring for both merging hadronic and merging hybrid stars)
and the correlation between binary mass and tidal deformability at the threshold for prompt black-hole formation.  
The strength of these signatures depend on details of the phase transition and the equation of state of quark matter. This implies that under certain circumstances it may be challenging to reveal or exclude the occurrence of quark deconfinement. This highlights the necessity of further research investigating more hybrid models and details of the various signals from neutron star mergers. 

The prospects to identify signals of quark confinement through the various features described above emphasizes the outstanding importance of future measurements and sensitivity upgrades of existing instruments. This includes in particular the ability to detect postmerger gravitational-wave emission in the kHz range but also electromagnetic follow-up observations to infer kilonova properties. 

\section{Epilogue}
\label{sec:1.6}

We want to take up the discussion of the Introduction that the simulations of core-collapse SN explosions and BNS mergers 
with model EOS are a unique tool to investigate the QCD phase diagram in the region of low temperatures and high baryon densities
($T\lesssim 60$ MeV at $1\lesssim n/n_0 \lesssim 5$) 
which is inaccessible to lattice QCD simulations and HIC experiments.
Not only that the detection of the above-described signals of a strong first-order phase transition in BNS mergers and/or supernovae would provide support for the existence of a CEP in the phase diagram that has so far been unsuccessfully been sought for in HIC experiments, there is theoretical evidence for a crossover transition at very low temperatures that suggests the existence of a second CEP
or even a crossover-all-over situation.  
This arises from the observation \cite{Hatsuda:2006ps,Abuki:2010jq} that a coexistence of chiral symmetry breaking and diquark condensation occurs at low temperatures due to the $U_A(1)$ 
anomaly-generating triangle diagram which, after Fierz transformation mixes diquark and meson condensates. 
This effect realizes the concept of quark-hadron continuity \cite{Schafer:1998ef} in a crossover transition. 
Support for such a picture comes also from recent progress in NS phenomenology.
\\
We are currently witnessing a paradigm change in the interpretation of mass and radius measurements of pulsars that is induced by the observation that from the multi-messenger analysis of typical-mass neutron star radii with $R_{1.4~M_\odot}=11.7^{+0.86}_{-0.81}$ km 
\cite{Dietrich:2020efo} (see also \cite{Capano:2019eae}) and the recent NICER radius measurement $R_{2.0~M_\odot}=13.7^{+2.6}_{-1.5}$ km \cite{Miller:2021qha} (see also \cite{Riley:2021pdl}) follows that $R_{2.0~M_\odot}\gtrsim R_{1.4~M_\odot}$.
The description of such a behaviour as solution of the Tolman-Oppenheimer-Volkoff equations requires a soft-stiff transition in the EOS at densities $n\lesssim 2 n_0$, just before the hyperon onset. 
This transition could be the hadron-to-quark matter transition.
In recent descriptions one joins a standard nuclear EOS with a 
constant speed of sound (CSS) model for the high-density phase either by a first-order phase transition (with a vanishing speed of sound
in the mixed phase \cite{Somasundaram:2021ljr}) or by directly matching the nuclear and quark matter squared speed of sound $c_s^2$ at a certain transition density $n_{\rm tr}$ without a density jump, thus mimicking a crossover transition \cite{Drischler:2020fvz}.
The best phenomenological description fulfilling simultaneously 
the constraints on both radii $R_{2.0~M_\odot}$
and $R_{1.4~M_\odot}$ is obtained in this simple picture by 
$n_{\rm tr}\sim 0.5~n_0$ and $c_s^2\sim 0.5$. 
We would like to remark that a CSS model with 
$c_s^2= 0.45 \dots 0.54$ provides an excellent fit to a microscopic nonlocal chiral quark model of the Nambu--Jona-Lasinio (NJL) type 
with diquark condensation (color superconductivity) and repulsive vector meson mean field 
\cite{Antic:2021zbn,Contrera:2022tqh}.
A direct, one-zone interpolation scheme between the safely known soft nuclear matter EoS (up to about $1.1~n_0$ as in \citep{Hebeler:2013nza})
and the suitably chosen stiff quark matter EoS (e.g., from a NJL model with coupling to a repulsive vector meson mean field) was pioneered in the works of \cite{Masuda:2012kf,Masuda:2012ed}.
Such a phase transition construction can be understood as a shortcut
for three physical effects as ingredients: 
\begin{enumerate}
\item[(i)] a stiffening of the nuclear matter EoS $P_H(\mu)$ due to the repulsive quark Pauli blocking effect between nucleons \cite{Ropke:1986qs,Blaschke:2020qrs} which can be effectively accounted for with a nucleonic excluded volume (see, e.g., \cite{Alvarez-Castillo:2016oln}), 
\item[(ii)] a strong reduction of the quark matter pressure $P_Q(\mu)$ at low chemical potentials due to confining forces (which result then in a good crossing of curves at $\mu=\mu_c$ allowing for the Maxwell construction $P_H(\mu_c)=P_Q(\mu_c)$ of a first-order phase transition), and
\item[(iii)] a mixed phase construction (e.g., by a parabolic interpolation   \cite{Ayriyan:2017nby}) that mimics the effects of finite-size structures (pasta phases) in the quark-hadron coexistence region. 
\end{enumerate}
For more details on the physics background, \citep[see][and references therein]{Ayriyan:2021prr,Baym:2018}.
With this microphysical basis behind the interpolation approach, a
two-zone interpolation construction for the hadron-to-quark matter has been developed in \cite{Ayriyan:2021prr}, where at the matching point $\mu_c$ situated between $\mu_H$ and $\mu_Q$, one can choose the condition of continuous density ($\Delta n=0$, crossover) or a finite density jump ($\Delta n \neq 0$, first-order transition). 
In concluding this epilogue, we want to give an outlook to the generalization of this two-zone interpolation scheme to finite temperatures (and arbitrary isospin densities) as a necessary prerequisite for investigating the consequences of these recent developments in the interpretation of neutron star phenomenology at zero temperature to simulations of supernova explosions and of binary neutron star merger events. 
The goal is to model the general class of hybrid EoS that corresponds to a phase diagram which has not only one critical endpoint (CEP) at high temperatures which marks the change from a first-order to a crossover transition regime, but also second CEP at low temperatures that arises from the competition and mixing between dynamical chiral symmetry breaking and color superconductivity.
Within the finite-temperature generalization of the two-zone interpolation scheme, this can be achieved by defining the function 
$\Delta n[\mu_c(T)]$ along the matching line $\mu_c(T)$ between the hadron-like and the quark-like interpolation zone in the phase diagram.  
The function $\Delta n[\mu_c(T)]$ encodes the position of the CEPs
$T_{\rm CEP,1}$ and $T_{\rm CEP,2}$ where $\Delta n=0$ as well as the strength of thee first-order transition between these points where  
$\Delta n \neq 0$.
\\
It is the aim of our ongoing research to investigate the dependence of the above described signals of a strong phase transition in supernova explosions and binary neutron star mergers on the detailed structure of the QCD phase diagram at low temperatures and high baryon densities and thus to be prepared for interpreting the possible observation of signals from such events in the near future.

\subsection*{Acknowledgements}
This work was supported by the Polish National Science Centre (NCN) under grant No. 2019/33/B/ST9/03059
(D.B.) and No. 2020/37/B/ST9/00691 (T.F.).
A.B. acknowledges support by the European Research Council (ERC) under the European Union's Horizon 2020 research and innovation programme under grant agreement No. 759253, by Deutsche Forschungsgemeinschaft (DFG, German Research Foundation) - Project-ID 279384907 - SFB 1245, by DFG - Project-ID 138713538 - SFB 881 ("The Milky Way System", subproject A10) and by the State of Hesse within the Cluster Project ELEMENTS.
The work was performed within a project that has received funding from the European Union's Horizon 2020 research and innovation program under grant agreement STRONG - 2020 - No. 824093.








\end{document}